\documentclass[aps,prapplied,reprint,superscriptaddress]{revtex4-1}

\usepackage{color}
\usepackage{graphicx}
\usepackage{dcolumn}
\usepackage{bm}

\begin{document}


\title{Room Temperature Electrically Detected Nuclear Spin Coherence of NV Centers in Diamond}


\author{H. Morishita}
\email[]{h-mori@scl.kyoto-u.ac.jp}
\affiliation{Institute for Chemical Research, Kyoto University, Gokasho, Uji, Kyoto 611-0011, Japan}
\affiliation{CREST, Japan Science and Technology Agency, Kawaguchi, Saitama 332-0012, Japan}

\author{S. Kobayashi}
\affiliation{Institute for Chemical Research, Kyoto University, Gokasho, Uji, Kyoto 611-0011, Japan}
\affiliation{CREST, Japan Science and Technology Agency, Kawaguchi, Saitama 332-0012, Japan}

\author{M. Fujiwara}
\affiliation{Institute for Chemical Research, Kyoto University, Gokasho, Uji, Kyoto 611-0011, Japan}
\affiliation{CREST, Japan Science and Technology Agency, Kawaguchi, Saitama 332-0012, Japan}

\author{H. Kato}
\affiliation{CREST, Japan Science and Technology Agency, Kawaguchi, Saitama 332-0012, Japan}
\affiliation{Energy Technology Research Institute, National Institute of Advanced Industrial Science and Technology (AIST), Tsukuba, Ibaraki 305-8568, Japan}

\author{T. Makino}
\affiliation{CREST, Japan Science and Technology Agency, Kawaguchi, Saitama 332-0012, Japan}
\affiliation{Energy Technology Research Institute, National Institute of Advanced Industrial Science and Technology (AIST), Tsukuba, Ibaraki 305-8568, Japan}

\author{S. Yamasaki}
\affiliation{CREST, Japan Science and Technology Agency, Kawaguchi, Saitama 332-0012, Japan}
\affiliation{Energy Technology Research Institute, National Institute of Advanced Industrial Science and Technology (AIST), Tsukuba, Ibaraki 305-8568, Japan}

\author{N. Mizuochi}
\email[]{mizuochi@scl.kyoto-u.ac.jp}
\affiliation{Institute for Chemical Research, Kyoto University, Gokasho, Uji, Kyoto 611-0011, Japan}
\affiliation{CREST, Japan Science and Technology Agency, Kawaguchi, Saitama 332-0012, Japan}


\date{\today}

\begin{abstract}
We demonstrate electrical detection of the $^{14}$N nuclear spin coherence of NV centers at room temperature. Nuclear spins are candidates for quantum memories in quantum-information devices and quantum sensors, and hence the electrical detection of nuclear spin coherence is essential to develop and integrate such quantum devices. In the present study, we used a pulsed electrically detected electron-nuclear double resonance technique to measure the Rabi oscillations and coherence time ($T_2$) of $^{14}$N nuclear spins in NV centers at room temperature. We observed $T_2 \approx$ 0.9 ms at room temperature. Our results will pave the way for the development of novel electron- and nuclear-spin-based diamond quantum devices.
\end{abstract}


\maketitle

\section{Introduction}
Nuclear spins in a semiconductor have a long coherence time ($T_2$) due to the good isolation from environmental noise~\cite{YusaNature05,MortonNature08, FuchsNatPhys11,MaurerSci12,PlaNature13,SaeediScience13,SigllitoNatNanotech17}. Therefore, they are candidates for quantum memories in quantum-information devices and quantum sensors. Using nuclear spins (e.g., nitrogen and carbon) in diamond for quantum memories, highly sensitive magnetic sensors~\cite{ZaiserNC16,MatsuzakiPRA16,PfenderNatCom17}, quantum repeaters~\cite{YangNatPhoto16}, quantum registers~\cite{NeumannScience08,WaldherrNature14}, etc., have been demonstrated at room temperature. In these demonstrations, the detection of nuclear spin coherence is essential. Nuclear spin coherence can be detected via the electron spins of nitrogen-vacancy (NV) centers, which also have a long $T_2$ at room temperature~\cite{BalasubramanianNM09,MizuochiPRB09}. NV electron spins can be detected by optical techniques~\cite{Doherty13} and electrical techniques~\cite{BourgeoisNComm15,HrubeschPRL2017,GulkaPRAppl17}. The electrical technique is an important technology for developing and integrating quantum devices. Furthermore, a theoretical model predicts that its detection sensitivity is approximately three times higher than that of the optical technique~\cite{HrubeschPRL2017}. While the electrical detection of the electron spin coherence of an ensemble of NV centers at room temperature has been demonstrated~\cite{HrubeschPRL2017,GulkaPRAppl17}, the direct electrical detection of nuclear spin coherence remains challenging. This is because nuclear spins are well isolated from the environment and the interactions between nuclear spins and current is very weak. Thus, we focus on an electrically detected electron-nuclear double resonance (EDENDOR) technique to demonstrate room-temperature electrical detection of nuclear spin coherence.

The first EDENDOR measurement was demonstrated by Stich and collaborators for phosphorus (P) donors in silicon at 4.2 K~\cite{StichAPL96}. After this demonstration, two other groups independently demonstrated pulse EDENDOR measurements of P-donors in silicon~\cite{McCameySci10,HoehnePRL11,McCameyPRB12,DreherPRL12}. They measured Rabi oscillations and $T_2$ of P-donor nuclear spins from 3.5 to 5 K. Furthermore, pulsed EDENDOR measurements of proton nuclear spins in organic semiconductors have been demonstrated~\cite{MalissaSci14}. Proton nuclear spin resonances have been measured at room temperature, but Rabi and $T_2$ measurements of proton nuclear spins have not been reported yet. To the best of our knowledge, there have been no demonstrations of room-temperature electrical detection of nuclear spin coherence in diamond or any other materials.

\begin{figure}
	\includegraphics[width=8cm,clip]{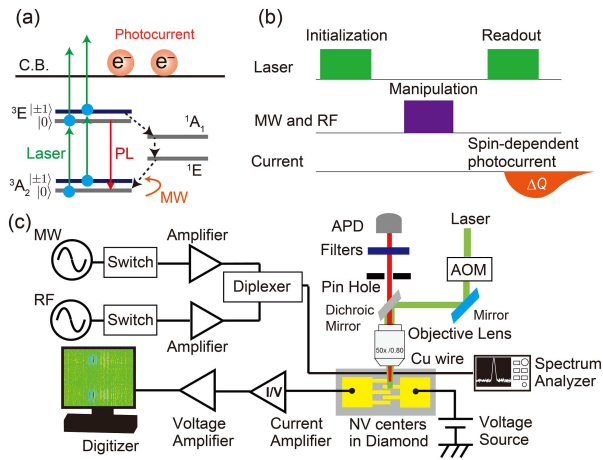}
	\caption{ (color online) 
		(a) Schematic of the EDENDOR measurements of NV centers in diamond. 
		(b) Process for the EDENDOR measurements. 
		(c) Schematic of the self-built EDENDOR spectrometer.
	}
	\label{figSchematics}
\end{figure}

The EDENDOR signals of $^{14}$N nuclear-spin coherence are observed by measuring the change in the electrically detected electron-spin echo (ESE) intensity of the NV centers. Here, the technique of electrical detection of magnetic resonance signals is called electrically detected magnetic resonance (EDMR). The EDMR of the NV centers measures the photocurrent change due to electron-spin resonances of the NV centers~\cite{BourgeoisNComm15,HrubeschPRL2017,GulkaPRAppl17}. The photocurrent can be generated under illumination by a 532-nm laser via a two-photon ionization process, depicted in Fig.~\ref{figSchematics}(a). The figure shows that the $\left|\pm 1\right>$ NV electron spin at the $^3E$ state has a transition probability to the long-lived ($\sim$ 220 ns) metastable $^1E$ states, where $\left|m_s\right>$ describes an NV electron spin. This causes the photocurrent to decrease due to a magnetic resonance transition from $\left|0\right>$ to $\left|\pm 1\right>$ after the optical initialization to $\left|0\right>$~\cite{BourgeoisNComm15}. Based on this mechanism, this study measures the electron and nuclear spin coherences with the pulse sequence depicted in Fig.~\ref{figSchematics}(b). The details of the pulse sequence are as follows. 1) A 532-nm laser is used to initialize the NV electron spins to $\left|0\right>$. 2) The electron and/or nuclear spins are manipulated by microwave (MW) and/or radio-frequency (RF), respectively. 3) The final laser is used to generate the photocurrent via two-photon ionization, which contains the EDMR and EDENDOR signals. They are observed as a transient response, which is depicted in Fig.~\ref{figSchematics}(b). Hence, this study defines the EDMR and EDENDOR intensities ($\Delta Q$) as the change in the photocurrent integrated by the time during which the photocurrent is changing~\cite{BoehmePRB03}.

\section{Method}
\subsection{Sample Preparation}
In this study, we used an ensemble of NV centers in a P-doped $n$-type diamond layer. This sample has two characteristics: 1) a negative charge state of the NV center (NV$^-$) in the P-doped $n$-type diamond that is stable even under 532-nm laser illumination~\cite{DoiPRB2016} and 2) the highly P-doped $n$-type diamond has high electrical conductivity~\cite{KatoAPL05}. The ensemble of NV centers in the $n$-type diamond was prepared via the following processes. A P-doped $n$-type diamond layer (10-$\mu$m thick) was synthesized on a type IIa (001) diamond substrate by chemical vapor deposition (CVD)~\cite{KatoAPL05}. The $n$-type diamond layer has a P-donor concentration of $\sim$ 10$^{18}$ cm$^{-3}$. The ensemble of NV centers was made by $^{14}$N$^+$-ion implantation (dose: 1 $\times$ 10$^{15}$ cm$^{-2}$) with a kinetic energy of 350 keV, followed by annealing at 1000 $^\circ$C for 1 hour under vacuum. We estimated the concentration of NV centers in the detection volume of the self-built EDENDOR spectrometer. The detection volume of the confocal microscope used as a laser illumination unit in the EDENDOR spectrometer can be estimated as 2 $\times$ 10$^{-12}$ cm$^3$~\cite{SM}. When the fluorescence from the NV center is proportional to the concentration of NV centers, we can estimate the concentration of NV centers of 1 $\times$ 10$^{15}$ cm$^{-3}$ (2 $\times$ 10$^3$ NV centers in the spot size of the focused laser beam). After generating the ensemble of NV centers, interdigital contacts with $\sim$ 2-$\mu$m gaps were fabricated on the $n$-type diamond layer to detect photocurrent changes, by the following three steps: 1) electron-beam lithography, 2) deposition of Ti(30 nm)/Pt(30 nm)/Au(100 nm) multilayers, and 3) annealing at 420 $^\circ$C for 30 min under an argon atmosphere.  Details of the electrical characteristics of the electrical contacts are given in Supplemental Material.

\subsection{Self-built EDENDOR Spectrometer}
Figure~\ref{figSchematics}(c) shows a self-built EDENDOR spectrometer~\cite{SM}. This spectrometer consists of 1) a laser illumination unit of a confocal microscope with a 532-nm laser to generate a photocurrent from an ensemble of NV centers, 2) a microwave (MW) and radio-frequency (RF) irradiation unit to manipulate the NV electron and $^{14}$N nuclear spins, and 3) a photocurrent detection unit. Using the laser illumination unit, the laser illuminates the ensemble of NV centers, focused by an objective lens with an NA of 0.8. Then, a photocurrent is generated from the NV centers. The NV centers also emit photons under the laser illumination, which are detected by an avalanche photodiode (APD) after passing through a pinhole and a filter (633-nm long pass filter). In this study, the APD is used to fix the position of the illumination spot in a place between the electrical contacts by monitoring the photons from the NV centers. MW and RF are combined with a frequency diplexer and irradiated to the NV centers with a $\sim$ 50-$\mu$m copper wire. The irradiated MW and RF frequencies and powers are measured by a spectrum analyzer during the EDMR and EDENDOR measurements. The change of photocurrent is measured under the application of a constant voltage of 8 V. The change of photocurrent is converted into a change of voltage by a current amplifier. Then, it can be measured by a digitizer on a personal computer after amplification by a voltage amplifier. A phase cycling technique is applied to subtract the artifact noises due to on- and off-resonant MW and RF contributions and laser-power fluctuations from the EDMR and EDENDOR signals~\cite{MalissaSci14, SM, pEPRBook,HoehnePRB13}. Note that the phases of the MW pulses are indicated by $\pm x$ on the MW pulses shown in Figs.~\ref{figENDOR}, \ref{figNRabi}, and \ref{fignEcho}. 

\section{EDMR OF ENSEMBLE OF NV CENTERS}
\subsection{Pulsed EDMR Spectrum of Ensemble of NV Centers}
\begin{figure}
	\includegraphics[width=7cm,clip]{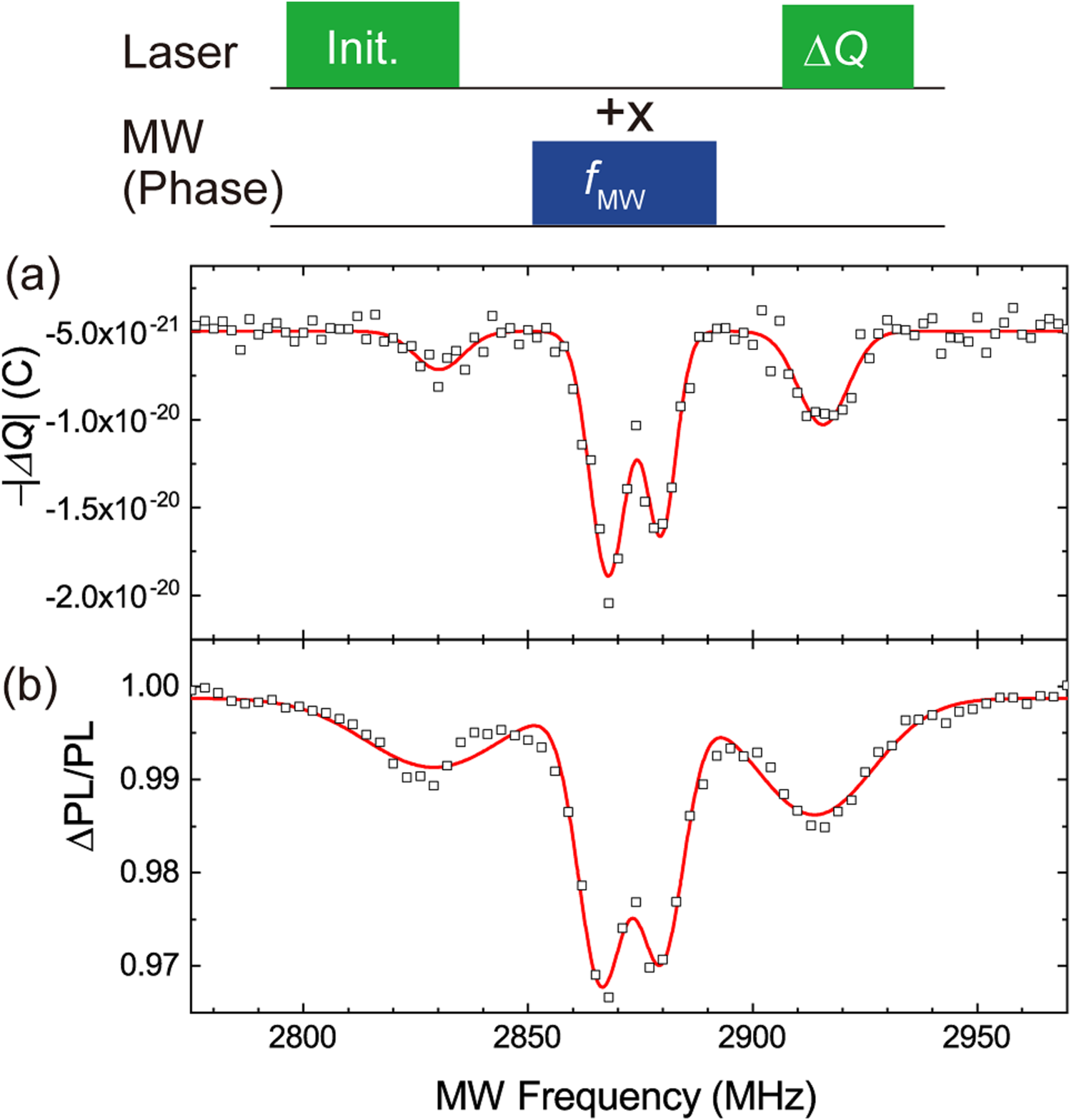}
	\caption{(color online)
		(Top) Pulse sequence. 
		(a) pEDMR spectrum. 
		(b) pODMR spectrum.
	}
	\label{figpEDMR}
\end{figure}

Initially, we performed a pulsed EDMR (pEDMR) measurement with the pulse sequence depicted in the top of Fig.~\ref{figpEDMR}. We used 30 mW of laser power and an input MW power of 10 mW at a static magnetic field of $\sim$ 10 G approximately along the [111] direction of the diamond crystal. Figure~\ref{figpEDMR}(a) shows the photocurrent change as a function of MW frequency, showing that four signals appeared. To analyze the observed spectrum, we measured the pulsed ODMR (pODMR) spectrum shown in Fig.~\ref{figpEDMR}(b) with the same conditions as the pEDMR measurements. The pEDMR spectrum has different linewidths than the pODMR spectrum. This may be due to the input laser power. While the pEDMR measurements require the illumination of the 30-mW laser to generate the photocurrent, the illumination of the 30-mW laser is strong enough to make broad linewidths in the pODMR signals. However, both spectra have the same resonant frequencies, and hence the NV electron spin resonance signals are observed via the electrical technique.

\subsection{Rabi Oscillations of NV Electron Spins }
\begin{figure}
	\includegraphics[width=8.5cm,clip]{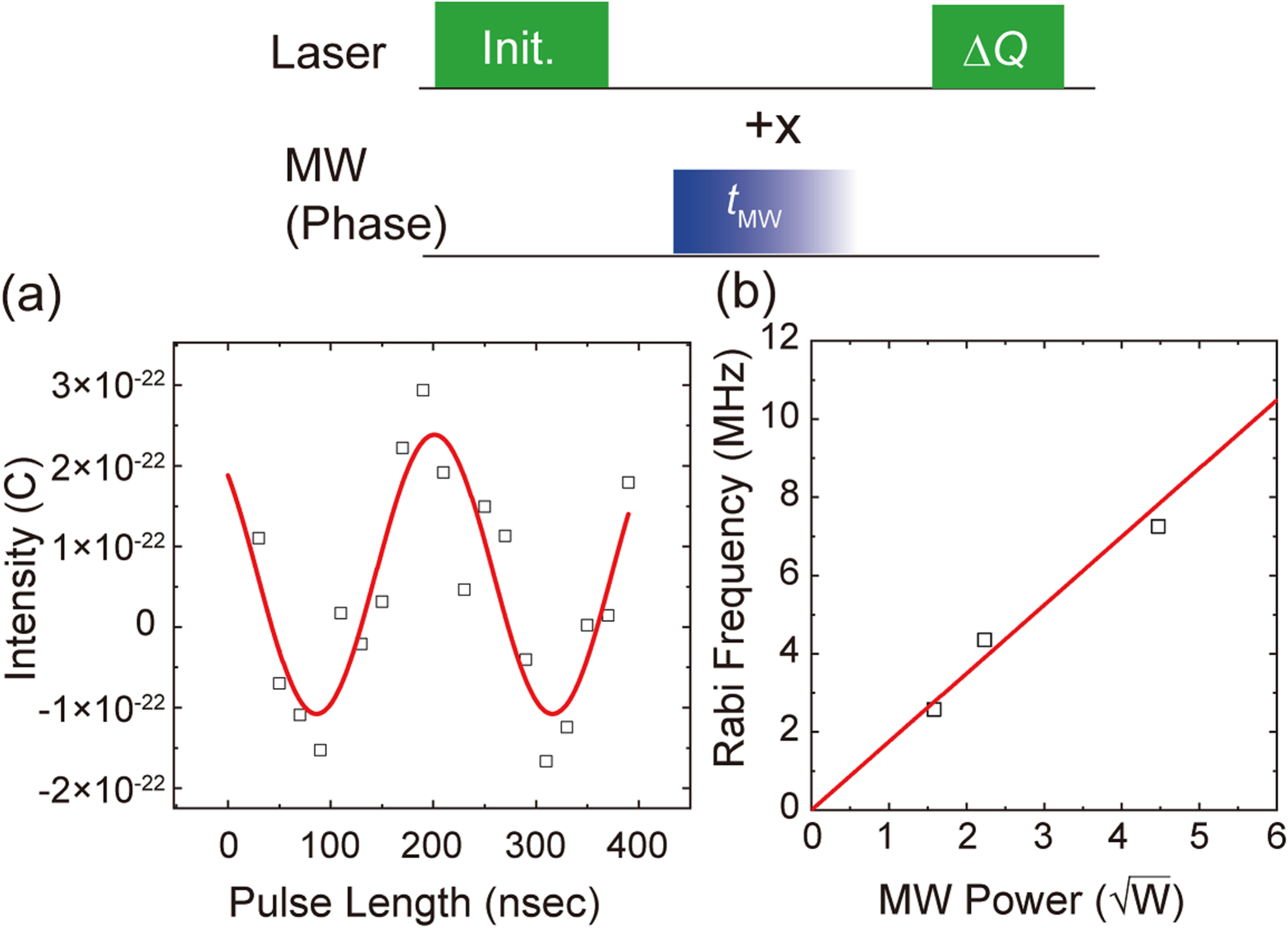}
	\caption{(color online) 
		(Top) Pulse sequence. 
		(a) Electrically detected electron-spin Rabi oscillation. 
		(b) Rabi frequencies as a function of the square root of MW power. 
	}
	\label{figRabi}
\end{figure}

The top of Fig.~\ref{figRabi} shows the pulse sequence for the electron-spin Rabi oscillations. After the initialization of the NV electron spins by the first laser pulse, we measured $\Delta Q$ under the application of the last laser as a function of the length of the resonant MW pulse ($t_\mathrm{MW}$). Here, we set the MW frequency to 2916 MHz, corresponding to the transition between $\left|0\right>$ and $\left|+1\right>$, and its input power to 5 W. The result is shown at the bottom of Fig.~\ref{figRabi}(a). The solid line in Fig.~\ref{figRabi}(a) shows the curve fitting result with a sinusoidal curve. We observed an oscillation frequency of $\sim$ 4.4 MHz. Next, we measured the oscillation frequencies with three different input MW powers, as depicted in Fig.~\ref{figRabi}(b). The figure plots the oscillation frequencies as a function of the square root of the input MW power. The observed data can be fitted by a linear function fixing the intercept at zero, and hence we observed Rabi oscillations of NV electron spins and the Rabi frequencies could be controlled by changing the input MW power.

\section{EDENDOR OF $^{14}$N NUCLEAR SPINS}
\subsection{Davies ENDOR Spectrum}
\begin{figure}
	\includegraphics[width=7cm,clip]{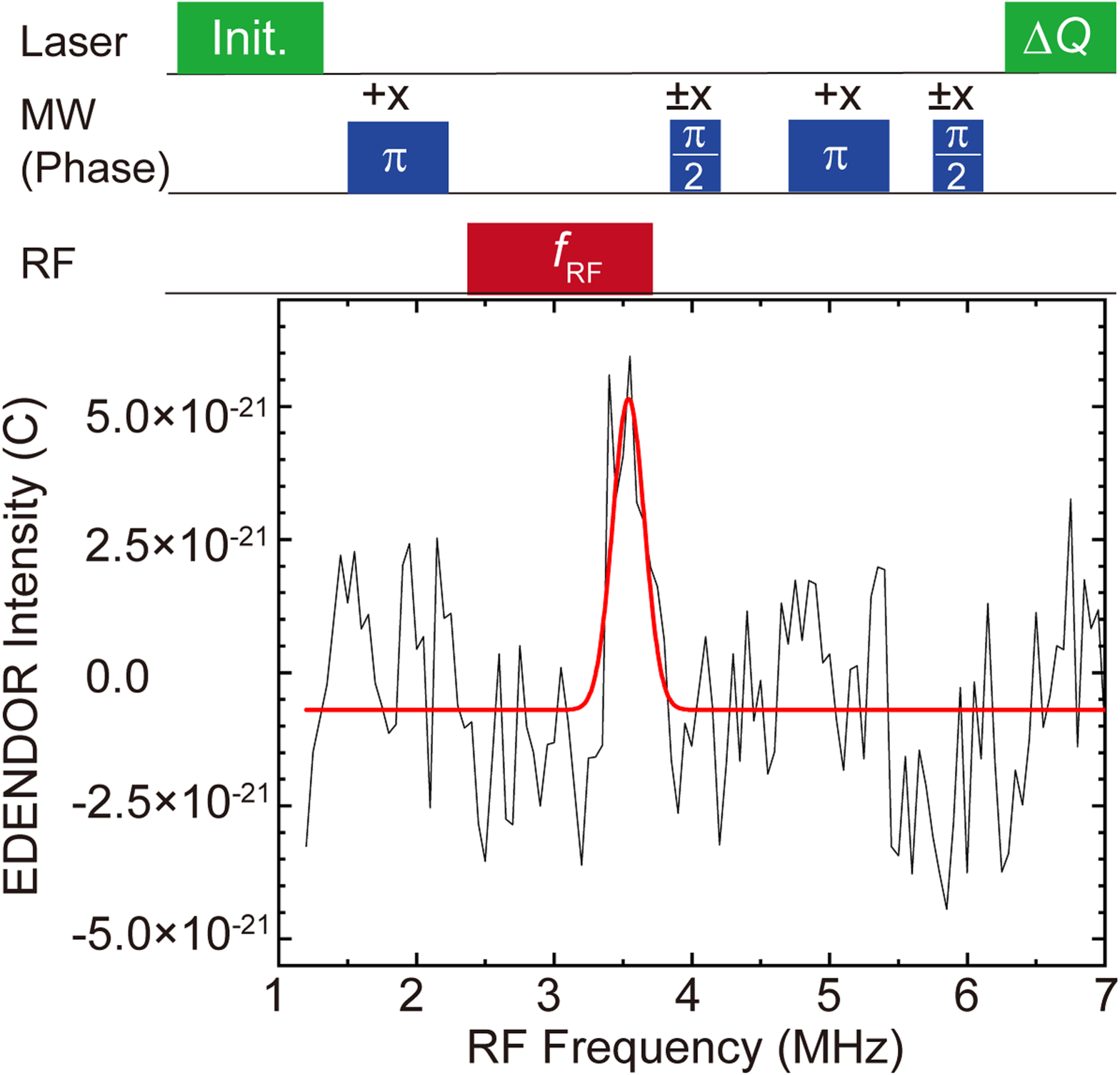}
	\caption{(color online) 
		Pulse sequence (top) and result (bottom) of an electrically detected Davies ENDOR spectrum.
	}
	\label{figENDOR}
\end{figure}

To measure the nuclear magnetic resonance of $^{14}$N nuclear spins, we used the Davies ENDOR technique. This enables the detection of a nuclear magnetic resonance (NMR) signal via a change in the ESE intensity based on the polarization transfer between the electron and nuclear transitions using the pulse sequence depicted in the top of Fig.~\ref{figENDOR}~\cite{HoehnePRL11,SM,pEPRBook}. First, an MW $\pi$-pulse is applied to the transition between $\left|0\right>$ to $\left|+1\right>$ after illumination by a pulsed laser. This $\pi$-pulse can generate hyperfine coupling between NV electron spins and $^{14}$N nuclear spins and the polarization between $\left|+1, 0\right>$ and $\left|+1, +1\right>$, where $\left|m_s, m_I\right>$ are electron and nuclear spins, respectively~\cite{SM,HePRB93,FeltonPRB09,YavkinJMR16}. Then, the RF pulse is applied. Finally, $\Delta Q$ was measured by applying a Hahn echo sequence and the following laser pulse. We observed Davies ENDOR spectra by measuring $\Delta Q$ as a function of the irradiated RF frequency.

Setting the MW frequency to 2916 MHz, the input MW power to 5 W, and the input RF power to 5 W, we observed the spectrum shown in Fig.~\ref{figENDOR}. The observed data can be fitted with the Gaussian function shown by the solid red line in Fig.~\ref{figENDOR}. The curve fitting result shows that we observed a resonance frequency of 3.5 MHz. Using Eq. (S1) in the Supplemental Material, we can estimate that the observed resonance frequency corresponds to the transition between $\left|+1, 0\right>$ and $\left|+1, +1\right>$. In the following measurements of the Rabi oscillations and the coherence time of the $^{14}$N nuclear spins, we will use the resonance frequency of 3.5 MHz. 

\subsection{Rabi Oscillations of $^{14}$N Nuclear Spins}
\begin{figure}
	\includegraphics[width=8cm,clip]{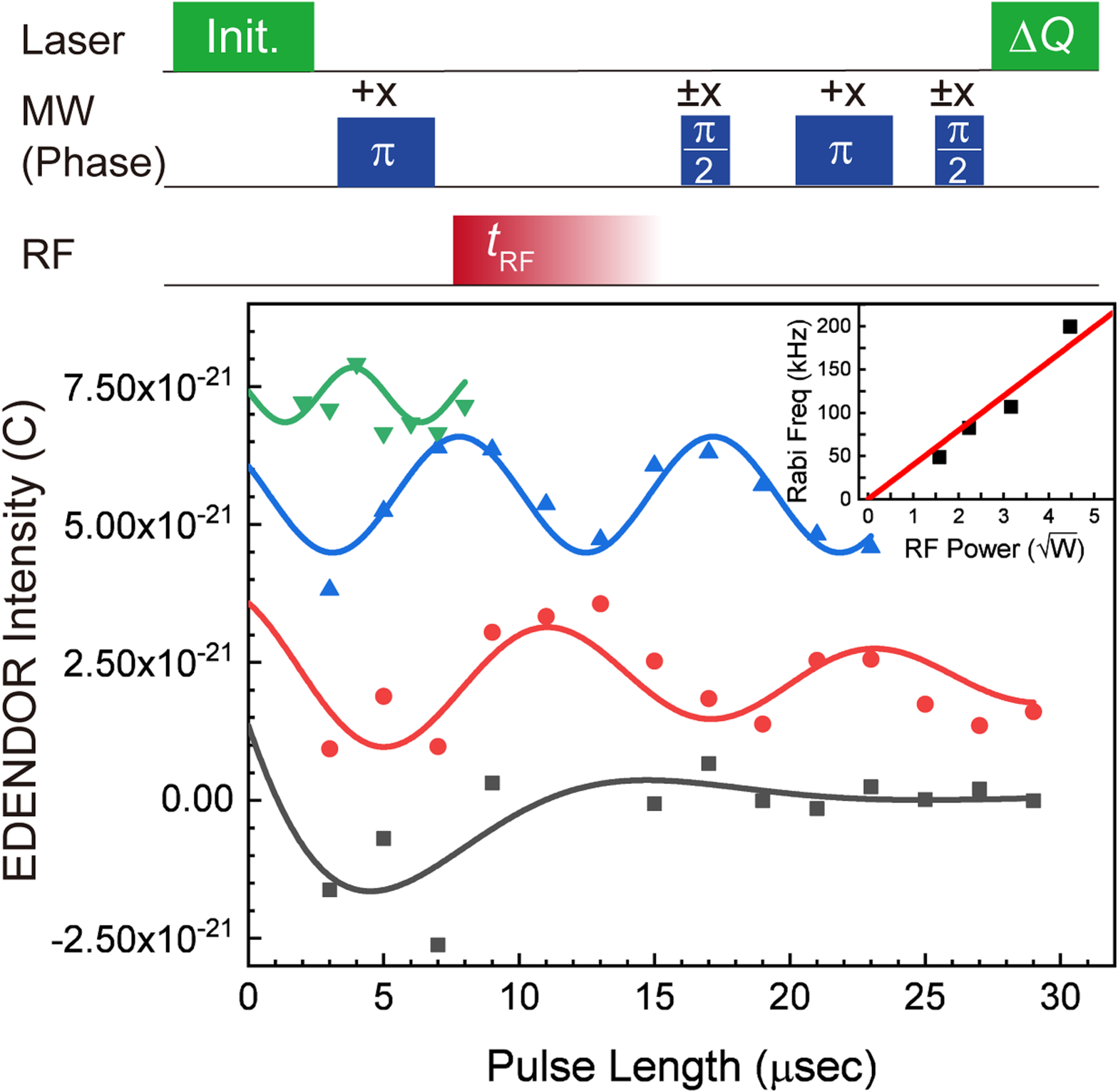}
	\caption{(color online) 
		Pulse sequence (top) and result (bottom) of an electrically detected nuclear Rabi oscillation. 
		(Inset) Rabi frequencies as a function of the square root of RF power. 
	}
	\label{figNRabi}
\end{figure}

The top of Fig.~\ref{figNRabi} shows the pulse sequence for the measurements of the $^{14}$N nuclear-spin Rabi oscillations. The sequence shows that we measured $\Delta Q$ as a function of the length of the RF pulse. Here, we set the MW frequency to 2916 MHz, the input MW power to 5 W, and the RF frequency to 3.5 MHz. The results of the electrically detected nuclear-spin Rabi measurements with four different input RF powers are shown in the bottom of Fig.~\ref{figNRabi}. The red, black, blue, and green points correspond to the results with input RF powers of $\sim$ 2.5, 5, 10, and 20 W, respectively. The observed oscillations are fitted by sinusoidal curves, which are shown as solid lines in the bottom of Fig.~\ref{figNRabi}. The oscillation frequencies observed by the curve fittings are plotted as a function of the square root of the input RF power in the inset of Fig.~\ref{figNRabi}. The plots are fitted well by a linear function with the intercept at zero, as shown by the solid line, which certifies that the observed oscillations correspond to Rabi oscillations between $\left|+1, 0\right>$ and $\left|+1, +1\right>$. Hence, the Rabi oscillations of the $^{14}$N nuclear spins can be observed with EDENDOR at room temperature.

\subsection{Echo Decay of $^{14}$N Nuclear Spins}
\begin{figure}
	\includegraphics[width=8cm,clip]{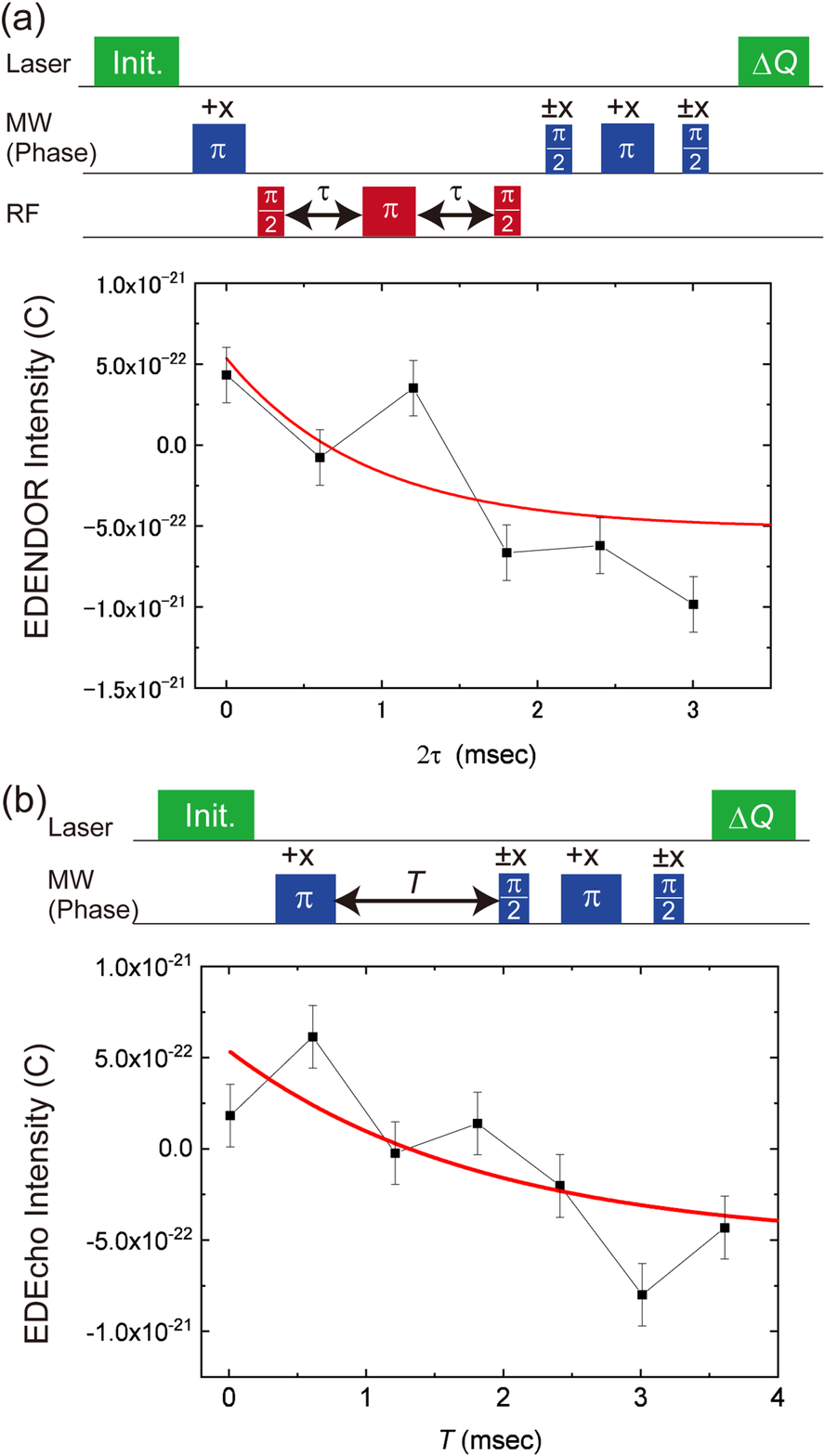}
	\caption{(color online) 
		(a) Pulse sequence (top) and result (bottom) of an electrically detected $T_2^{(n)}$ measurement of the $^{14}$N nuclear spins in NV centers. 
		(b) Pulse sequence (top) and result and electrically detected $T_1^{(e)}$ measurement of the NV electron spins (bottom).
	}
	\label{fignEcho}
\end{figure}

The top of Fig.~\ref{fignEcho}(a) shows the pulse sequence for a $^{14}$N nuclear spin ($T_2^{(n)}$) measurement. It can also be modified based on the Davies EDENDOR sequence to measure $T_2^{(n)}$~\cite{McCameyPRB12,DreherPRL12,HoehnePRB13}. A nuclear-spin Hahn echo sequence was added between the first MW $\pi$- and second MW $\pi$/2-pulse, allowing the nuclear spin echo intensity to be measured via the change in the ESE intensity. We set the input RF power to 10 W and the other experimental conditions were the same as those for electrically detected nuclear spin Rabi oscillations. Then, we measured $\Delta Q$ as a change of the ESE intensity as a function of the freely evolving time of 2$\tau$. The bottom of Fig.~\ref{fignEcho}(a) shows the result of the $T_2^{(n)}$ measurement. The experimentally observed plots are fitted with an exponential curve described by the solid line in Fig.~\ref{fignEcho}(a) fixing the echo amplitude at half the Rabi amplitude. Consequently, $T_2^{(n)} \approx $ 0.9 (5) ms. Hence, we successfully observed $T_2^{(n)}$ with the EDENDOR technique at room temperature.

\section{DISCUSSION}
We here discuss the results of the $T_2^{(n)}$ measurement with the modified Davies ENDOR sequence. After the application of the first MW $\pi$-pulse in the sequence of the $T_2^{(n)}$ measurement depicted in Fig.~\ref{fignEcho}(a), the NV electron spins relax back to their thermal equilibrium with a time constant defined as the longitudinal relaxation time $T_1^{(e)}$. Thus, the observed $T_2^{(n)}$ may be limited by $T_1^{(e)}$ in our experiments. We measured $T_1^{(e)}$ with the EDMR technique. The pulse sequence for a $T_1^{(e)}$ measurement is shown in the top of Fig.~\ref{fignEcho}(b). It is the same as the sequence for $T_2^{(n)}$ measurements, without the train of RF pulses. Using the same experimental conditions as those for the $T_2^{(n)}$ measurement, except for the irradiation by the RF pulses, we measured $\Delta Q$ as the change in the ESE intensity as a function of $T$, which is the interval between the first MW $\pi$- and second MW $\pi$/2-pulses. In the bottom of Fig.~\ref{fignEcho}(b), the experimentally observed data points were fitted with the exponential curve described by the solid line. This gave $T_1^{(e)} \approx $ 1.8(6) ms, which is slightly shorter than the reported $T_1^{(e)}$~\cite{NeumannScience08}, which may be due to the additional decay rate into $\left|0\right>$ due to laser leakage through an acousto-optic modulator (AOM) and/or noises in the magnetic and electric fields from P-donors~\cite{MyersPRL17,AriyaratneNC18} and/or from a dark electrical current in the absence of the laser illumination. Myers et al.~\cite{MyersPRL17} showed that the electrical field from the diamond surface noise influences the $T_1^{(e)}$ of the NV electron spins. Since the electron carriers released from the P-donors and the positive charge of the P-donors after the release generates the electric field, the $T_1^{(e)}$ of the NV electron spins may become short in the highly P-doped diamond layer. Moreover, the $T_1^{(e)}$ measurement was performed in the presence of a dark current~\cite{SM}. The dark current produces magnetic fluctuations~\cite{AriyaratneNC18}, which may shorten $T_1^{(e)}$.

Finally, we compare the observed $T_1^{(e)}$ with $T_2^{(n)}$. When the $T_2^{(n)}$ decay is affected only by $T_1^{(e)}$, the theory expects $T_2^{(n)} = \frac{3}{2}T_1^{(e)}$~\cite{PfenderNatCom17}. However, the observed $T_2^{(n)}$ is slightly shorter than $T_1^{(e)}$. When we consider the influence of laser leakage through the AOM accompanied by additional decay rate into $\left|0\right>$, $T_2^{(n)}$ might become shorter than $\frac{3}{2}T_1^{(e)}$, as discussed in Ref.~\cite{PfenderNatCom17}. Thus, precise control of the concentration of P-donors and the upgrade of the EDENDOR spectrometer are important in order to observe long $T_1^{(e)}$ and $T_2^{(n)}$ by the EDENDOR technique.

\section{SUMMARY}
We have demonstrated electrical detection of the Rabi oscillations and $T_2^{(n)}$ of the $^{14}$N nuclear spin coherence of an ensemble of NV centers in diamond at room temperature. Using a self-built EDENDOR spectrometer, we observed a signal at $\sim$ 3.5 MHz, which is the ENDOR frequency of the $^{14}$N nuclear spins. This frequency was used to demonstrate electrically detected nuclear-spin Rabi oscillations and $T_2^{(n)}$ measurements of the $^{14}$N nuclear spins at room temperature. We observed $T_2^{(n)} \approx$ 0.9 ms. This study should contribute to the development of future electron- and nuclear-spin based diamond quantum devices.

\section{ACKNOWLEDGMENTS}
We wish to acknowledge T. Ono and T. Moriyama for their support with the fabrication and evaluation of the electrical contacts. This work was supported by CREST (No. JPMJCR1333), KAKENHI (No. 15H05868), Kyoto University Nano Technology Hub, and the Collaborative Research Program of the Institute for Chemical Research, Kyoto University (Grant No. 2017-78). HM acknowledges support from KAKENHI (No. 16K17848 and No. 19H02546), Murata Science Foundation, and Iketani Science and Technology Foundation. 
%

\newpage{}
\section*{Supplemental Material}
\renewcommand{\thefigure}{S\arabic{figure}}
\setcounter{figure}{0}
\subsection*{NV Center Coupled to a $^{14}$N Nuclear Spin}
The Hamiltonian for an NV electron spin coupled with a $^{14}$N nuclear spin under a static magnetic field ($B_0$)~\cite{SDoherty13} is:
\begin{eqnarray}
\displaystyle{\cal{H}} &=&  D_{gs}\left[S_z^2 -\frac{1}{3} S(S+1) \right] + g_e\mu_BB_0S_z  \nonumber \\
&& + g_n\mu_nB_0I_z + A_\parallel S_z I_z + A_\perp \left(S_x I_x + S_y  I_y\right) \nonumber \\
&&+ P \left[I_z^2  -\frac{1}{3} I(I+1) \right]. 
\end{eqnarray}
The first and second terms are described as the zero-field splitting and Zeeman interaction of NV electron spins, respectively, where $D_{gs} \sim$   2.87 GHz, $S_z$, $g_e$ $\sim$ 2.003, and $\mu_B$ are the zero-field splitting parameter, $z$ component of the NV electron spin ($S$), $g$-factor of the NV electron spin, and Bohr magnetron, respectively. The third term is described as the Zeeman interaction of $^{14}$N nuclear spin, where $g_n \sim$  0.404 and $\mu_n$ are the $g$-factor of the $^{14}$N nuclear spin and nuclear magneton, respectively~\cite{SMackRMP50}. The fourth and fifth terms are described as the axial ($A_\parallel \sim -2.1$ MHz ) and non-axial hyperfine interactions ($A_\perp  \sim -2.7$ MHz) with the $^{14}$N nuclear spin, respectively~\cite{SFeltonPRB09}. The sixth term is described as the quadrupole interaction of the $^{14}$N nuclear spin ( $P \sim -5.0$  MHz)~\cite{SFeltonPRB09}. $I_x$, $I_y$, and $I_z$ are $x$, $y$, and $z$ components of the $^{14}$N nuclear spin ($I$), respectively. 

\subsection*{Self-built EDENDOR Spectrometer}
\begin{figure}
	\includegraphics[width=8cm,clip]{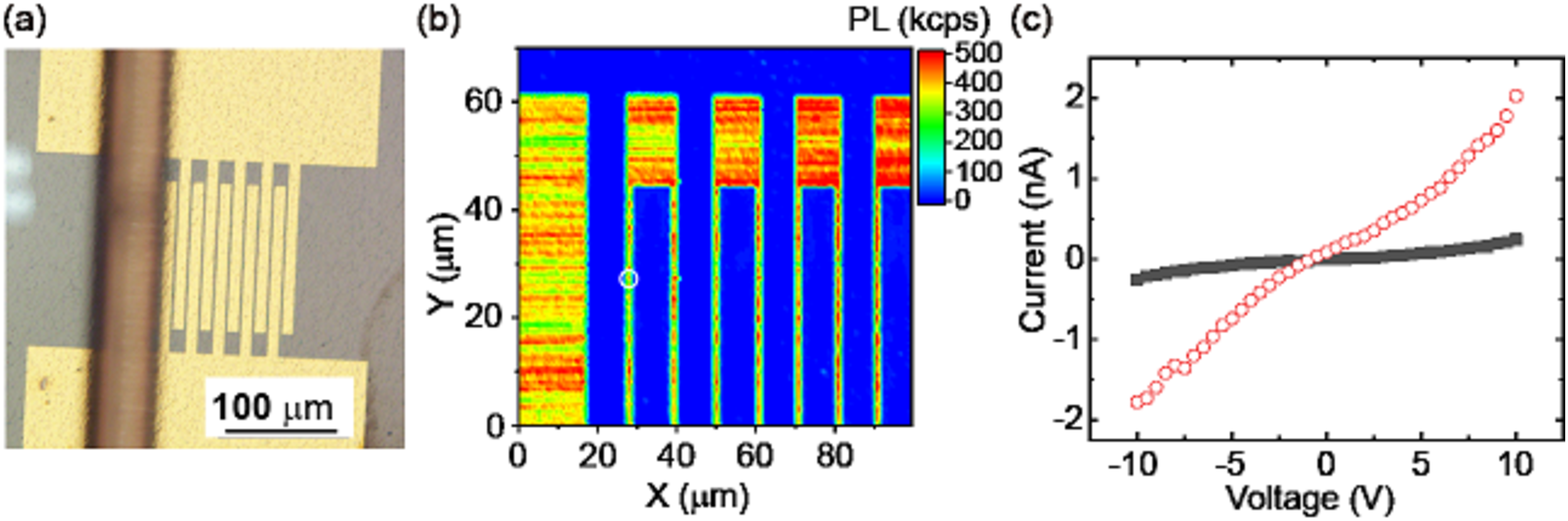}
	\caption{ 
		(color online)
		(a) Photograph of electrical contacts with MW antenna. 
		(b) PL laser-scan image of electrical contacts. 
		(c) IV characteristics of electrical contacts. The hollow and filled points show the IV characteristics with and without laser illumination, respectively.
	}
	\label{fig:Optical}
\end{figure}

The self-built EDENDOR spectrometer consists of the following three units: 1) laser illumination unit, 2) microwave (MW)- and radio-frequency (RF)-irradiation unit, and 3) photocurrent detection unit, depicted in Fig. 1(c) of the main text. A confocal laser microscope with a 532-nm laser acts as the laser illumination unit. The 532-nm laser was pulsed by an acousto-optic modulator (AOM). After the pulsed laser is reflected by a dichroic mirror, it illuminates the ensemble of NV centers focused by an objective lens with an NA of 0.8. For an objective lens with an NA of 0.8 and a laser with a wavelength of 532 nm, we can estimate the detection volume of our confocal laser microscope as 2 $\times$ 10$^{-12}$ cm$^3$~\cite{SCFMBook}. Using the laser illumination unit, the photocurrent from the NV centers is generated under the laser illumination. Moreover, photons emitted from the NV centers were detected by an avalanche photodiode (APD) after passing through a pinhole with a diameter of $\sim$ 30 $\mu$m and two filters (an 835-nm short-pass filter and a 633-nm long-pass filter). In this study, the APD was used to fix the position of the illumination spot in the place depicted in the white circle in Fig.~\ref{fig:Optical}(b). In the MW- and RF-irradiation unit, MW and RF generated by two high-frequency oscillators are pulsed by frequency switches. After amplification of the pulsed MW and RF with MW and RF amplifiers, they are combined with a frequency diplexer. They are then irradiated to the NV centers by a copper wire with a diameter of $\sim$ 50 $\mu$m. We used a spectrum analyzer to measure the irradiated MW and RF frequencies and powers during the EDMR and EDENDOR measurements. We measured  the change of photocurrent under the application of a constant voltage (SRS SIM928) with the photocurrent detection unit. In the photocurrent detection unit, the change of photocurrents is converted to a change of voltage by a current amplifier (FEMTO DHPCA-100). Then, the change of voltage can be measured by a digitizer (Gage Razor CSE1621) on a personal computer after amplification with a voltage amplifier (FEMTO DHPVA-201). 

\subsection*{Evaluation of Electrical Contacts }
Figures~\ref{fig:Optical}(a) and \ref{fig:Optical}(b) show a photograph and PL laser-scan image with a laser power of 15 $\mu$W of the electrical contacts for the EDMR and EDENDOR measurements. We set the laser power to 30 mW and fixed the laser spot to the position depicted in the white circle in Fig.~\ref{fig:Optical}(b), and measured the current-voltage characteristics of the electrical contacts with and without laser illumination. The hollow and filled points in Fig.~\ref{fig:Optical}(c) show the current-voltage characteristics with and without laser illumination, respectively. It shows that the dark current, which is the current in the absence of laser illumination, flows in the diamond and a photocurrent is generated under laser illumination. 

\begin{figure}
	\includegraphics[width=8cm,clip]{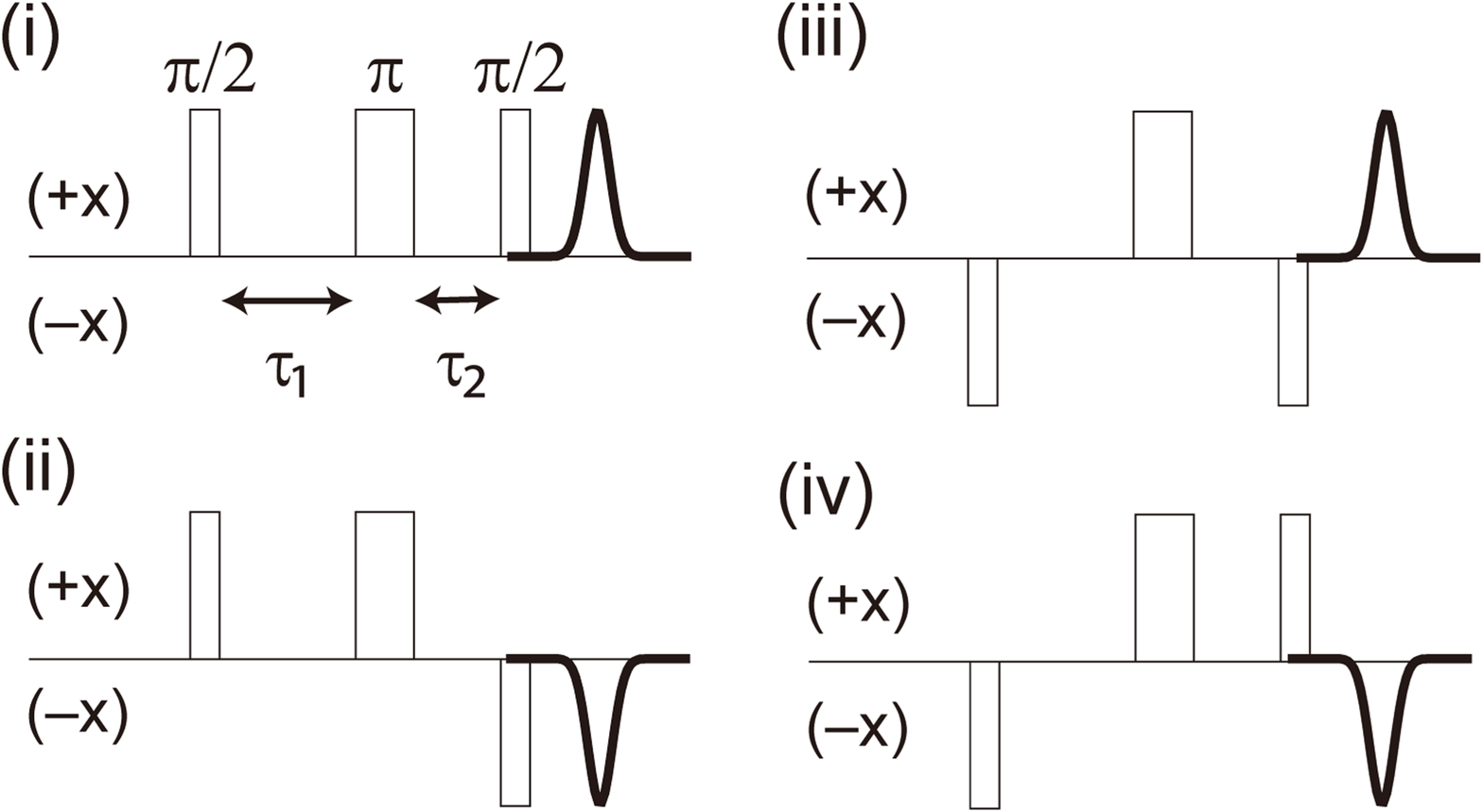}
	\caption{ 
		(color online)
		Electron-spin Hahn echo measurements with four different phase cycling configurations.
	}
	\label{fig:PhaseCycling}
\end{figure}

\begin{figure}
	\includegraphics[width=8cm,clip]{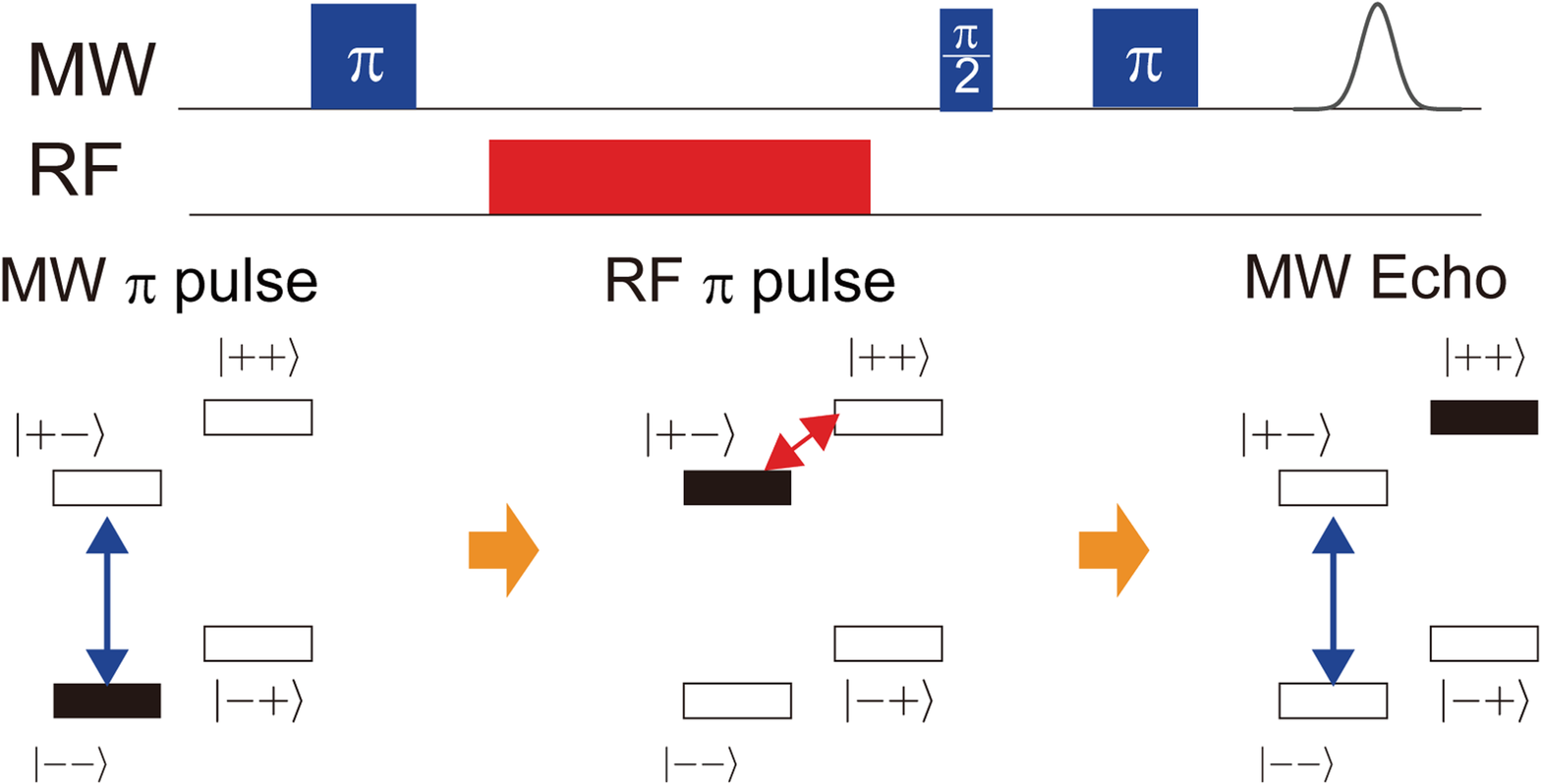}
	\caption{ 
		(color online)
		Davies ENDOR pulse sequence (Top) and polarization transfers with the above pulse sequence in the quantum system of an electron spin ($S$ = 1/2) and nuclear spin ($I$ = 1/2) (Bottom).
	}
	\label{fig:DaviesENDOR}
\end{figure}

\subsection*{Phase Cycling}
A phase cycling technique is used to subtract the on- and off-resonant MW and RF contributions and fluctuations of laser power to the magnetic resonance signals. The phases of the MW pulses are indicated by $\pm x$ on the MW pulses in the pulse sequences depicted in Figs. 2, 3, 4, 5, and 6 of the main text. To explain this phase cycling technique, we discuss the polarities of Hahn-echo signals with the four different phase cycling configurations depicted in Fig.~\ref{fig:PhaseCycling}. We observe positive echo signals in cases (i) and (iii) of Fig.~\ref{fig:PhaseCycling} and negative echo signals in case (ii) and (iv) of Fig.~\ref{fig:PhaseCycling}~\cite{SpEPRBook, SHoehnePRB13,SMalissaSci14}. On the other hand, the polarities of the MW and RF currents due to the irradiation of on- and off-resonant MW and RF fields and photocurrent noise induced by the laser power fluctuation do not change with the above phase cycling configurations. Thus, we observe just the magnetic resonance signals by adding the echo signals with the sequences (i) and (iii) to the sum of the magnetic resonance signals and subtracting the echo signals with the sequences (ii) and (iv) from one~\cite{SpEPRBook, SHoehnePRB13,SMalissaSci14}.

\subsection*{Davies ENDOR}

We consider the quantum system of an electron spin ($S$ = 1/2) coupled with a nuclear spin ($I$ = 1/2). The top of Fig.~\ref{fig:DaviesENDOR} shows the pulse sequence of the Davies ENDOR~\cite{SpEPRBook}, and the bottom of Fig.~\ref{fig:DaviesENDOR} shows the energy levels of the system, where $\left| \pm, \pm \right>$ represent the electron and nuclear spins, respectively. After the system is initialized to $\left| -- \right>$, depicted by the filled rectangles, the MW-$\pi$ pulse inverts the polarization of the transition between $\left|--\right>$ and $\left|+-\right>$ (bottom left of Fig.~\ref{fig:DaviesENDOR}). Next, the resonant RF-$\pi$ pulse is applied to the transition between $\left|+-\right>$ and $\left|++\right>$. Then, its polarization is inverted (bottom center of Fig.~\ref{fig:DaviesENDOR}). Finally, we measured the echo of the transition between $\left|--\right>$ to $\left|+-\right>$ (bottom right of Fig.~\ref{fig:DaviesENDOR}). In this situation, we do not observe any echo signals. Such a change of the echo signal means that a nuclear-magnetic-resonance transition between $\left|+-\right>$ and $\left|++\right>$ occurs. Therefore, we can observe the nuclear magnetic resonance signals with the Davies ENDOR sequence.
 

\begin{thebibliography}{36}%
	\makeatletter
	\providecommand \@ifxundefined [1]{%
		\@ifx{#1\undefined}
	}%
	\providecommand \@ifnum [1]{%
		\ifnum #1\expandafter \@firstoftwo
		\else \expandafter \@secondoftwo
		\fi
	}%
	\providecommand \@ifx [1]{%
		\ifx #1\expandafter \@firstoftwo
		\else \expandafter \@secondoftwo
		\fi
	}%
	\providecommand \natexlab [1]{#1}%
	\providecommand \enquote  [1]{``#1''}%
	\providecommand \bibnamefont  [1]{#1}%
	\providecommand \bibfnamefont [1]{#1}%
	\providecommand \citenamefont [1]{#1}%
	\providecommand \href@noop [0]{\@secondoftwo}%
	\providecommand \href [0]{\begingroup \@sanitize@url \@href}%
	\providecommand \@href[1]{\@@startlink{#1}\@@href}%
	\providecommand \@@href[1]{\endgroup#1\@@endlink}%
	\providecommand \@sanitize@url [0]{\catcode `\\12\catcode `\$12\catcode
		`\&12\catcode `\#12\catcode `\^12\catcode `\_12\catcode `\%12\relax}%
	\providecommand \@@startlink[1]{}%
	\providecommand \@@endlink[0]{}%
	\providecommand \url  [0]{\begingroup\@sanitize@url \@url }%
	\providecommand \@url [1]{\endgroup\@href {#1}{\urlprefix }}%
	\providecommand \urlprefix  [0]{URL }%
	\providecommand \Eprint [0]{\href }%
	\providecommand \doibase [0]{https://doi.org/}%
	\providecommand \selectlanguage [0]{\@gobble}%
	\providecommand \bibinfo  [0]{\@secondoftwo}%
	\providecommand \bibfield  [0]{\@secondoftwo}%
	\providecommand \translation [1]{[#1]}%
	\providecommand \BibitemOpen [0]{}%
	\providecommand \bibitemStop [0]{}%
	\providecommand \bibitemNoStop [0]{.\EOS\space}%
	\providecommand \EOS [0]{\spacefactor3000\relax}%
	\providecommand \BibitemShut  [1]{\csname bibitem#1\endcsname}%
	\let\auto@bib@innerbib\@empty
	\bibitem [{\citenamefont {Yusa}\ \emph {et~al.}(2005)\citenamefont {Yusa},
		\citenamefont {Muraki}, \citenamefont {Takashina}, \citenamefont
		{Hashimoto},\ and\ \citenamefont {Hirayama}}]{YusaNature05}%
	\BibitemOpen
	\bibfield  {author} {\bibinfo {author} {\bibfnamefont {G.}~\bibnamefont
			{Yusa}}, \bibinfo {author} {\bibfnamefont {K.}~\bibnamefont {Muraki}},
		\bibinfo {author} {\bibfnamefont {K.}~\bibnamefont {Takashina}}, \bibinfo
		{author} {\bibfnamefont {K.}~\bibnamefont {Hashimoto}},\ and\ \bibinfo
		{author} {\bibfnamefont {Y.}~\bibnamefont {Hirayama}},\ }\href@noop {}
	{\bibfield  {journal} {\bibinfo  {journal} {Nature (London)}\ }\textbf
		{\bibinfo {volume} {434}},\ \bibinfo {pages} {1001} (\bibinfo {year}
		{2005})}\BibitemShut {NoStop}%
	\bibitem [{\citenamefont {Morton}\ \emph {et~al.}(2008)\citenamefont {Morton},
		\citenamefont {Tyryshkin}, \citenamefont {Brown}, \citenamefont {Shankar},
		\citenamefont {Lovett}, \citenamefont {Ardavan}, \citenamefont {Schenkel},
		\citenamefont {Haller}, \citenamefont {Ager},\ and\ \citenamefont
		{Lyon}}]{MortonNature08}%
	\BibitemOpen
	\bibfield  {author} {\bibinfo {author} {\bibfnamefont {J.~J.~L.}\
			\bibnamefont {Morton}}, \bibinfo {author} {\bibfnamefont {A.~M.}\
			\bibnamefont {Tyryshkin}}, \bibinfo {author} {\bibfnamefont {R.~M.}\
			\bibnamefont {Brown}}, \bibinfo {author} {\bibfnamefont {S.}~\bibnamefont
			{Shankar}}, \bibinfo {author} {\bibfnamefont {B.~W.}\ \bibnamefont {Lovett}},
		\bibinfo {author} {\bibfnamefont {A.}~\bibnamefont {Ardavan}}, \bibinfo
		{author} {\bibfnamefont {T.}~\bibnamefont {Schenkel}}, \bibinfo {author}
		{\bibfnamefont {E.~E.}\ \bibnamefont {Haller}}, \bibinfo {author}
		{\bibfnamefont {J.~W.}\ \bibnamefont {Ager}},\ and\ \bibinfo {author}
		{\bibfnamefont {S.~A.}\ \bibnamefont {Lyon}},\ }\href@noop {} {\bibfield
		{journal} {\bibinfo  {journal} {Nature (London)}\ }\textbf {\bibinfo {volume}
			{455}},\ \bibinfo {pages} {1085} (\bibinfo {year} {2008})}\BibitemShut
	{NoStop}%
	\bibitem [{\citenamefont {Fuchs}\ \emph {et~al.}(2011)\citenamefont {Fuchs},
		\citenamefont {Burkard}, \citenamefont {Klimov},\ and\ \citenamefont
		{Awschalom}}]{FuchsNatPhys11}%
	\BibitemOpen
	\bibfield  {author} {\bibinfo {author} {\bibfnamefont {G.~D.}\ \bibnamefont
			{Fuchs}}, \bibinfo {author} {\bibfnamefont {G.}~\bibnamefont {Burkard}},
		\bibinfo {author} {\bibfnamefont {P.~V.}\ \bibnamefont {Klimov}},\ and\
		\bibinfo {author} {\bibfnamefont {D.~D.}\ \bibnamefont {Awschalom}},\ }\href
	{https://doi.org/10.1038/nphys2026} {\bibfield  {journal} {\bibinfo
			{journal} {Nat. Phys.}\ }\textbf {\bibinfo {volume} {7}},\ \bibinfo {pages}
		{789} (\bibinfo {year} {2011})}\BibitemShut {NoStop}%
	\bibitem [{\citenamefont {Maurer}\ \emph {et~al.}(2012)\citenamefont {Maurer},
		\citenamefont {Kucsko}, \citenamefont {Latta}, \citenamefont {Jiang},
		\citenamefont {Yao}, \citenamefont {Bennett}, \citenamefont {Pastawski},
		\citenamefont {Hunger}, \citenamefont {Chisholm}, \citenamefont {Markham},
		\citenamefont {Twitchen}, \citenamefont {Cirac},\ and\ \citenamefont
		{Lukin}}]{MaurerSci12}%
	\BibitemOpen
	\bibfield  {author} {\bibinfo {author} {\bibfnamefont {P.~C.}\ \bibnamefont
			{Maurer}}, \bibinfo {author} {\bibfnamefont {G.}~\bibnamefont {Kucsko}},
		\bibinfo {author} {\bibfnamefont {C.}~\bibnamefont {Latta}}, \bibinfo
		{author} {\bibfnamefont {L.}~\bibnamefont {Jiang}}, \bibinfo {author}
		{\bibfnamefont {N.~Y.}\ \bibnamefont {Yao}}, \bibinfo {author} {\bibfnamefont
			{S.~D.}\ \bibnamefont {Bennett}}, \bibinfo {author} {\bibfnamefont
			{F.}~\bibnamefont {Pastawski}}, \bibinfo {author} {\bibfnamefont
			{D.}~\bibnamefont {Hunger}}, \bibinfo {author} {\bibfnamefont
			{N.}~\bibnamefont {Chisholm}}, \bibinfo {author} {\bibfnamefont
			{M.}~\bibnamefont {Markham}}, \bibinfo {author} {\bibfnamefont {D.~J.}\
			\bibnamefont {Twitchen}}, \bibinfo {author} {\bibfnamefont {J.~I.}\
			\bibnamefont {Cirac}},\ and\ \bibinfo {author} {\bibfnamefont {M.~D.}\
			\bibnamefont {Lukin}},\ }\href@noop {} {\bibfield  {journal} {\bibinfo
			{journal} {Science}\ }\textbf {\bibinfo {volume} {336}},\ \bibinfo {pages}
		{1283} (\bibinfo {year} {2012})}\BibitemShut {NoStop}%
	\bibitem [{\citenamefont {Pla}\ \emph {et~al.}(2013)\citenamefont {Pla},
		\citenamefont {Tan}, \citenamefont {Dehollain}, \citenamefont {Lim},
		\citenamefont {Morton}, \citenamefont {Zwanenburg}, \citenamefont {Jamieson},
		\citenamefont {Dzurak},\ and\ \citenamefont {Morello}}]{PlaNature13}%
	\BibitemOpen
	\bibfield  {author} {\bibinfo {author} {\bibfnamefont {J.~J.}\ \bibnamefont
			{Pla}}, \bibinfo {author} {\bibfnamefont {K.~Y.}\ \bibnamefont {Tan}},
		\bibinfo {author} {\bibfnamefont {J.~P.}\ \bibnamefont {Dehollain}}, \bibinfo
		{author} {\bibfnamefont {W.~H.}\ \bibnamefont {Lim}}, \bibinfo {author}
		{\bibfnamefont {J.~J.~L.}\ \bibnamefont {Morton}}, \bibinfo {author}
		{\bibfnamefont {F.~A.}\ \bibnamefont {Zwanenburg}}, \bibinfo {author}
		{\bibfnamefont {D.~N.}\ \bibnamefont {Jamieson}}, \bibinfo {author}
		{\bibfnamefont {A.~S.}\ \bibnamefont {Dzurak}},\ and\ \bibinfo {author}
		{\bibfnamefont {A.}~\bibnamefont {Morello}},\ }\href@noop {} {\bibfield
		{journal} {\bibinfo  {journal} {Nature (London)}\ }\textbf {\bibinfo {volume}
			{496}},\ \bibinfo {pages} {334} (\bibinfo {year} {2013})}\BibitemShut
	{NoStop}%
	\bibitem [{\citenamefont {Saeedi}\ \emph {et~al.}(2013)\citenamefont {Saeedi},
		\citenamefont {Simmons}, \citenamefont {Salvail}, \citenamefont {Dluhy},
		\citenamefont {Riemann}, \citenamefont {Abrosimov}, \citenamefont {Becker},
		\citenamefont {Pohl}, \citenamefont {Morton},\ and\ \citenamefont
		{Thewalt}}]{SaeediScience13}%
	\BibitemOpen
	\bibfield  {author} {\bibinfo {author} {\bibfnamefont {K.}~\bibnamefont
			{Saeedi}}, \bibinfo {author} {\bibfnamefont {S.}~\bibnamefont {Simmons}},
		\bibinfo {author} {\bibfnamefont {J.~Z.}\ \bibnamefont {Salvail}}, \bibinfo
		{author} {\bibfnamefont {P.}~\bibnamefont {Dluhy}}, \bibinfo {author}
		{\bibfnamefont {H.}~\bibnamefont {Riemann}}, \bibinfo {author} {\bibfnamefont
			{N.~V.}\ \bibnamefont {Abrosimov}}, \bibinfo {author} {\bibfnamefont
			{P.}~\bibnamefont {Becker}}, \bibinfo {author} {\bibfnamefont {H.-J.}\
			\bibnamefont {Pohl}}, \bibinfo {author} {\bibfnamefont {J.~J.~L.}\
			\bibnamefont {Morton}},\ and\ \bibinfo {author} {\bibfnamefont {M.~L.~W.}\
			\bibnamefont {Thewalt}},\ }\href@noop {} {\bibfield  {journal} {\bibinfo
			{journal} {Science}\ }\textbf {\bibinfo {volume} {342}},\ \bibinfo {pages}
		{830} (\bibinfo {year} {2013})}\BibitemShut {NoStop}%
	\bibitem [{\citenamefont {Sigillito}\ \emph {et~al.}(2017)\citenamefont
		{Sigillito}, \citenamefont {Tyryshkin}, \citenamefont {Schenkel},
		\citenamefont {Houck},\ and\ \citenamefont {Lyon}}]{SigllitoNatNanotech17}%
	\BibitemOpen
	\bibfield  {author} {\bibinfo {author} {\bibfnamefont {A.~J.}\ \bibnamefont
			{Sigillito}}, \bibinfo {author} {\bibfnamefont {A.~M.}\ \bibnamefont
			{Tyryshkin}}, \bibinfo {author} {\bibfnamefont {T.}~\bibnamefont {Schenkel}},
		\bibinfo {author} {\bibfnamefont {A.~A.}\ \bibnamefont {Houck}},\ and\
		\bibinfo {author} {\bibfnamefont {S.~A.}\ \bibnamefont {Lyon}},\ }\href@noop
	{} {\bibfield  {journal} {\bibinfo  {journal} {Nat. Nanotech.}\ }\textbf
		{\bibinfo {volume} {12}},\ \bibinfo {pages} {958} (\bibinfo {year}
		{2017})}\BibitemShut {NoStop}%
	\bibitem [{\citenamefont {Zaiser}\ \emph {et~al.}(2016)\citenamefont {Zaiser},
		\citenamefont {Rendler}, \citenamefont {Jakobi}, \citenamefont {Wolf},
		\citenamefont {Lee}, \citenamefont {Wagner}, \citenamefont {Bergholm},
		\citenamefont {Schulte-Herbr\"{u}ggen}, \citenamefont {Neumann},\ and\
		\citenamefont {Wrachtrup}}]{ZaiserNC16}%
	\BibitemOpen
	\bibfield  {author} {\bibinfo {author} {\bibfnamefont {S.}~\bibnamefont
			{Zaiser}}, \bibinfo {author} {\bibfnamefont {T.}~\bibnamefont {Rendler}},
		\bibinfo {author} {\bibfnamefont {I.}~\bibnamefont {Jakobi}}, \bibinfo
		{author} {\bibfnamefont {T.}~\bibnamefont {Wolf}}, \bibinfo {author}
		{\bibfnamefont {S.-Y.}\ \bibnamefont {Lee}}, \bibinfo {author} {\bibfnamefont
			{S.}~\bibnamefont {Wagner}}, \bibinfo {author} {\bibfnamefont
			{V.}~\bibnamefont {Bergholm}}, \bibinfo {author} {\bibfnamefont
			{T.}~\bibnamefont {Schulte-Herbr\"{u}ggen}}, \bibinfo {author} {\bibfnamefont
			{P.}~\bibnamefont {Neumann}},\ and\ \bibinfo {author} {\bibfnamefont
			{J.}~\bibnamefont {Wrachtrup}},\ }\href@noop {} {\bibfield  {journal}
		{\bibinfo  {journal} {Nat. Commun.}\ }\textbf {\bibinfo {volume} {7}},\
		\bibinfo {pages} {12279} (\bibinfo {year} {2016})}\BibitemShut {NoStop}%
	\bibitem [{\citenamefont {Matsuzaki}\ \emph {et~al.}(2016)\citenamefont
		{Matsuzaki}, \citenamefont {Shimo-Oka}, \citenamefont {Tanaka}, \citenamefont
		{Tokura}, \citenamefont {Semba},\ and\ \citenamefont
		{Mizuochi}}]{MatsuzakiPRA16}%
	\BibitemOpen
	\bibfield  {author} {\bibinfo {author} {\bibfnamefont {Y.}~\bibnamefont
			{Matsuzaki}}, \bibinfo {author} {\bibfnamefont {T.}~\bibnamefont
			{Shimo-Oka}}, \bibinfo {author} {\bibfnamefont {H.}~\bibnamefont {Tanaka}},
		\bibinfo {author} {\bibfnamefont {Y.}~\bibnamefont {Tokura}}, \bibinfo
		{author} {\bibfnamefont {K.}~\bibnamefont {Semba}},\ and\ \bibinfo {author}
		{\bibfnamefont {N.}~\bibnamefont {Mizuochi}},\ }\href@noop {} {\bibfield
		{journal} {\bibinfo  {journal} {Phys. Rev. A}\ }\textbf {\bibinfo {volume}
			{94}},\ \bibinfo {pages} {052330} (\bibinfo {year} {2016})}\BibitemShut
	{NoStop}%
	\bibitem [{\citenamefont {Pfender}\ \emph {et~al.}(2017)\citenamefont
		{Pfender}, \citenamefont {Aslam}, \citenamefont {Sumiya}, \citenamefont
		{Onoda}, \citenamefont {Neumann}, \citenamefont {Isoya}, \citenamefont
		{Meriles},\ and\ \citenamefont {Wrachtrup}}]{PfenderNatCom17}%
	\BibitemOpen
	\bibfield  {author} {\bibinfo {author} {\bibfnamefont {M.}~\bibnamefont
			{Pfender}}, \bibinfo {author} {\bibfnamefont {N.}~\bibnamefont {Aslam}},
		\bibinfo {author} {\bibfnamefont {H.}~\bibnamefont {Sumiya}}, \bibinfo
		{author} {\bibfnamefont {S.}~\bibnamefont {Onoda}}, \bibinfo {author}
		{\bibfnamefont {P.}~\bibnamefont {Neumann}}, \bibinfo {author} {\bibfnamefont
			{J.}~\bibnamefont {Isoya}}, \bibinfo {author} {\bibfnamefont {C.~A.}\
			\bibnamefont {Meriles}},\ and\ \bibinfo {author} {\bibfnamefont
			{J.}~\bibnamefont {Wrachtrup}},\ }\href@noop {} {\bibfield  {journal}
		{\bibinfo  {journal} {Nat. Commun.}\ }\textbf {\bibinfo {volume} {8}},\
		\bibinfo {pages} {834} (\bibinfo {year} {2017})}\BibitemShut {NoStop}%
	\bibitem [{\citenamefont {Yang}\ \emph {et~al.}(2016)\citenamefont {Yang},
		\citenamefont {Wang}, \citenamefont {Rao}, \citenamefont {Hien~Tran},
		\citenamefont {Momenzadeh}, \citenamefont {Markham}, \citenamefont
		{Twitchen}, \citenamefont {Wang}, \citenamefont {Yang}, \citenamefont
		{St\"{o}hr}, \citenamefont {Neumann}, \citenamefont {Kosaka},\ and\
		\citenamefont {Wrachtrup}}]{YangNatPhoto16}%
	\BibitemOpen
	\bibfield  {author} {\bibinfo {author} {\bibfnamefont {S.}~\bibnamefont
			{Yang}}, \bibinfo {author} {\bibfnamefont {Y.}~\bibnamefont {Wang}}, \bibinfo
		{author} {\bibfnamefont {D.~D.~B.}\ \bibnamefont {Rao}}, \bibinfo {author}
		{\bibfnamefont {T.}~\bibnamefont {Hien~Tran}}, \bibinfo {author}
		{\bibfnamefont {A.~S.}\ \bibnamefont {Momenzadeh}}, \bibinfo {author}
		{\bibfnamefont {M.}~\bibnamefont {Markham}}, \bibinfo {author} {\bibfnamefont
			{D.~J.}\ \bibnamefont {Twitchen}}, \bibinfo {author} {\bibfnamefont
			{P.}~\bibnamefont {Wang}}, \bibinfo {author} {\bibfnamefont {W.}~\bibnamefont
			{Yang}}, \bibinfo {author} {\bibfnamefont {R.}~\bibnamefont {St\"{o}hr}},
		\bibinfo {author} {\bibfnamefont {P.}~\bibnamefont {Neumann}}, \bibinfo
		{author} {\bibfnamefont {H.}~\bibnamefont {Kosaka}},\ and\ \bibinfo {author}
		{\bibfnamefont {J.}~\bibnamefont {Wrachtrup}},\ }\href@noop {} {\bibfield
		{journal} {\bibinfo  {journal} {Nat. Photon.}\ }\textbf {\bibinfo {volume}
			{10}},\ \bibinfo {pages} {507} (\bibinfo {year} {2016})}\BibitemShut
	{NoStop}%
	\bibitem [{\citenamefont {Neumann}\ \emph {et~al.}(2008)\citenamefont
		{Neumann}, \citenamefont {Mizuochi}, \citenamefont {Rempp}, \citenamefont
		{Hemmer}, \citenamefont {Watanabe}, \citenamefont {Yamasaki}, \citenamefont
		{Jacques}, \citenamefont {Gaebel}, \citenamefont {Jelezko},\ and\
		\citenamefont {Wrachtrup}}]{NeumannScience08}%
	\BibitemOpen
	\bibfield  {author} {\bibinfo {author} {\bibfnamefont {P.}~\bibnamefont
			{Neumann}}, \bibinfo {author} {\bibfnamefont {N.}~\bibnamefont {Mizuochi}},
		\bibinfo {author} {\bibfnamefont {F.}~\bibnamefont {Rempp}}, \bibinfo
		{author} {\bibfnamefont {P.}~\bibnamefont {Hemmer}}, \bibinfo {author}
		{\bibfnamefont {H.}~\bibnamefont {Watanabe}}, \bibinfo {author}
		{\bibfnamefont {S.}~\bibnamefont {Yamasaki}}, \bibinfo {author}
		{\bibfnamefont {V.}~\bibnamefont {Jacques}}, \bibinfo {author} {\bibfnamefont
			{T.}~\bibnamefont {Gaebel}}, \bibinfo {author} {\bibfnamefont
			{F.}~\bibnamefont {Jelezko}},\ and\ \bibinfo {author} {\bibfnamefont
			{J.}~\bibnamefont {Wrachtrup}},\ }\href@noop {} {\bibfield  {journal}
		{\bibinfo  {journal} {Science}\ }\textbf {\bibinfo {volume} {320}},\ \bibinfo
		{pages} {1326} (\bibinfo {year} {2008})}\BibitemShut {NoStop}%
	\bibitem [{\citenamefont {Waldherr}\ \emph {et~al.}(2014)\citenamefont
		{Waldherr}, \citenamefont {Wang}, \citenamefont {Zaiser}, \citenamefont
		{Jamali}, \citenamefont {Schulte-Herbr\"{u}ggen}, \citenamefont {Abe},
		\citenamefont {Ohshima}, \citenamefont {Isoya}, \citenamefont {Du},
		\citenamefont {Neumann},\ and\ \citenamefont {Wrachtrup}}]{WaldherrNature14}%
	\BibitemOpen
	\bibfield  {author} {\bibinfo {author} {\bibfnamefont {G.}~\bibnamefont
			{Waldherr}}, \bibinfo {author} {\bibfnamefont {Y.}~\bibnamefont {Wang}},
		\bibinfo {author} {\bibfnamefont {S.}~\bibnamefont {Zaiser}}, \bibinfo
		{author} {\bibfnamefont {M.}~\bibnamefont {Jamali}}, \bibinfo {author}
		{\bibfnamefont {T.}~\bibnamefont {Schulte-Herbr\"{u}ggen}}, \bibinfo {author}
		{\bibfnamefont {H.}~\bibnamefont {Abe}}, \bibinfo {author} {\bibfnamefont
			{T.}~\bibnamefont {Ohshima}}, \bibinfo {author} {\bibfnamefont
			{J.}~\bibnamefont {Isoya}}, \bibinfo {author} {\bibfnamefont {J.~F.}\
			\bibnamefont {Du}}, \bibinfo {author} {\bibfnamefont {P.}~\bibnamefont
			{Neumann}},\ and\ \bibinfo {author} {\bibfnamefont {J.}~\bibnamefont
			{Wrachtrup}},\ }\href@noop {} {\bibfield  {journal} {\bibinfo  {journal}
			{Nature (London)}\ }\textbf {\bibinfo {volume} {506}},\ \bibinfo {pages}
		{204} (\bibinfo {year} {2014})}\BibitemShut {NoStop}%
	\bibitem [{\citenamefont {Balasubramanian}\ \emph {et~al.}(2009)\citenamefont
		{Balasubramanian}, \citenamefont {Neumann}, \citenamefont {Twitchen},
		\citenamefont {Markham}, \citenamefont {Kolesov}, \citenamefont {Mizuochi},
		\citenamefont {Isoya}, \citenamefont {Achard}, \citenamefont {Beck},
		\citenamefont {Tissler}, \citenamefont {Jacques}, \citenamefont {Hemmer},
		\citenamefont {Jelezko},\ and\ \citenamefont
		{Wrachtrup}}]{BalasubramanianNM09}%
	\BibitemOpen
	\bibfield  {author} {\bibinfo {author} {\bibfnamefont {G.}~\bibnamefont
			{Balasubramanian}}, \bibinfo {author} {\bibfnamefont {P.}~\bibnamefont
			{Neumann}}, \bibinfo {author} {\bibfnamefont {D.}~\bibnamefont {Twitchen}},
		\bibinfo {author} {\bibfnamefont {M.}~\bibnamefont {Markham}}, \bibinfo
		{author} {\bibfnamefont {R.}~\bibnamefont {Kolesov}}, \bibinfo {author}
		{\bibfnamefont {N.}~\bibnamefont {Mizuochi}}, \bibinfo {author}
		{\bibfnamefont {J.}~\bibnamefont {Isoya}}, \bibinfo {author} {\bibfnamefont
			{J.}~\bibnamefont {Achard}}, \bibinfo {author} {\bibfnamefont
			{J.}~\bibnamefont {Beck}}, \bibinfo {author} {\bibfnamefont {J.}~\bibnamefont
			{Tissler}}, \bibinfo {author} {\bibfnamefont {V.}~\bibnamefont {Jacques}},
		\bibinfo {author} {\bibfnamefont {P.~R.}\ \bibnamefont {Hemmer}}, \bibinfo
		{author} {\bibfnamefont {F.}~\bibnamefont {Jelezko}},\ and\ \bibinfo {author}
		{\bibfnamefont {J.}~\bibnamefont {Wrachtrup}},\ }\href@noop {} {\bibfield
		{journal} {\bibinfo  {journal} {Nat. Mater.}\ }\textbf {\bibinfo {volume}
			{8}},\ \bibinfo {pages} {383} (\bibinfo {year} {2009})}\BibitemShut {NoStop}%
	\bibitem [{\citenamefont {Mizuochi}\ \emph {et~al.}(2009)\citenamefont
		{Mizuochi}, \citenamefont {Neumann}, \citenamefont {Rempp}, \citenamefont
		{Beck}, \citenamefont {Jacques}, \citenamefont {Siyushev}, \citenamefont
		{Nakamura}, \citenamefont {Twitchen}, \citenamefont {Watanabe}, \citenamefont
		{Yamasaki}, \citenamefont {Jelezko},\ and\ \citenamefont
		{Wrachtrup}}]{MizuochiPRB09}%
	\BibitemOpen
	\bibfield  {author} {\bibinfo {author} {\bibfnamefont {N.}~\bibnamefont
			{Mizuochi}}, \bibinfo {author} {\bibfnamefont {P.}~\bibnamefont {Neumann}},
		\bibinfo {author} {\bibfnamefont {F.}~\bibnamefont {Rempp}}, \bibinfo
		{author} {\bibfnamefont {J.}~\bibnamefont {Beck}}, \bibinfo {author}
		{\bibfnamefont {V.}~\bibnamefont {Jacques}}, \bibinfo {author} {\bibfnamefont
			{P.}~\bibnamefont {Siyushev}}, \bibinfo {author} {\bibfnamefont
			{K.}~\bibnamefont {Nakamura}}, \bibinfo {author} {\bibfnamefont {D.~J.}\
			\bibnamefont {Twitchen}}, \bibinfo {author} {\bibfnamefont {H.}~\bibnamefont
			{Watanabe}}, \bibinfo {author} {\bibfnamefont {S.}~\bibnamefont {Yamasaki}},
		\bibinfo {author} {\bibfnamefont {F.}~\bibnamefont {Jelezko}},\ and\ \bibinfo
		{author} {\bibfnamefont {J.}~\bibnamefont {Wrachtrup}},\ }\href@noop {}
	{\bibfield  {journal} {\bibinfo  {journal} {Phys. Rev. B}\ }\textbf {\bibinfo
			{volume} {80}},\ \bibinfo {pages} {041201(R)} (\bibinfo {year}
		{2009})}\BibitemShut {NoStop}%
	\bibitem [{\citenamefont {Doherty}\ \emph {et~al.}(2013)\citenamefont
		{Doherty}, \citenamefont {Manson}, \citenamefont {Delaney}, \citenamefont
		{Jelezko}, \citenamefont {Wrachtrup},\ and\ \citenamefont
		{Hollenberg}}]{Doherty13}%
	\BibitemOpen
	\bibfield  {author} {\bibinfo {author} {\bibfnamefont {M.~W.}\ \bibnamefont
			{Doherty}}, \bibinfo {author} {\bibfnamefont {N.~B.}\ \bibnamefont {Manson}},
		\bibinfo {author} {\bibfnamefont {P.}~\bibnamefont {Delaney}}, \bibinfo
		{author} {\bibfnamefont {F.}~\bibnamefont {Jelezko}}, \bibinfo {author}
		{\bibfnamefont {J.}~\bibnamefont {Wrachtrup}},\ and\ \bibinfo {author}
		{\bibfnamefont {L.~C.~L.}\ \bibnamefont {Hollenberg}},\ }\href@noop {}
	{\bibfield  {journal} {\bibinfo  {journal} {Phys. Rep.}\ }\textbf {\bibinfo
			{volume} {528}},\ \bibinfo {pages} {1} (\bibinfo {year} {2013})}\BibitemShut
	{NoStop}%
	\bibitem [{\citenamefont {Bourgeois}\ \emph {et~al.}(2015)\citenamefont
		{Bourgeois}, \citenamefont {Jarmola}, \citenamefont {Siyushev}, \citenamefont
		{Gulka}, \citenamefont {Hruby}, \citenamefont {Jelezko}, \citenamefont
		{Budker},\ and\ \citenamefont {Nesladek}}]{BourgeoisNComm15}%
	\BibitemOpen
	\bibfield  {author} {\bibinfo {author} {\bibfnamefont {E.}~\bibnamefont
			{Bourgeois}}, \bibinfo {author} {\bibfnamefont {A.}~\bibnamefont {Jarmola}},
		\bibinfo {author} {\bibfnamefont {P.}~\bibnamefont {Siyushev}}, \bibinfo
		{author} {\bibfnamefont {M.}~\bibnamefont {Gulka}}, \bibinfo {author}
		{\bibfnamefont {J.}~\bibnamefont {Hruby}}, \bibinfo {author} {\bibfnamefont
			{F.}~\bibnamefont {Jelezko}}, \bibinfo {author} {\bibfnamefont
			{D.}~\bibnamefont {Budker}},\ and\ \bibinfo {author} {\bibfnamefont
			{M.}~\bibnamefont {Nesladek}},\ }\href@noop {} {\bibfield  {journal}
		{\bibinfo  {journal} {Nat. Commun.}\ }\textbf {\bibinfo {volume} {6}},\
		\bibinfo {pages} {8577} (\bibinfo {year} {2015})}\BibitemShut {NoStop}%
	\bibitem [{\citenamefont {Hrubesch}\ \emph {et~al.}(2017)\citenamefont
		{Hrubesch}, \citenamefont {Braunbeck}, \citenamefont {Stutzmann},
		\citenamefont {Reinhard},\ and\ \citenamefont {Brandt}}]{HrubeschPRL2017}%
	\BibitemOpen
	\bibfield  {author} {\bibinfo {author} {\bibfnamefont {F.~M.}\ \bibnamefont
			{Hrubesch}}, \bibinfo {author} {\bibfnamefont {G.}~\bibnamefont {Braunbeck}},
		\bibinfo {author} {\bibfnamefont {M.}~\bibnamefont {Stutzmann}}, \bibinfo
		{author} {\bibfnamefont {F.}~\bibnamefont {Reinhard}},\ and\ \bibinfo
		{author} {\bibfnamefont {M.~S.}\ \bibnamefont {Brandt}},\ }\href@noop {}
	{\bibfield  {journal} {\bibinfo  {journal} {Phys. Rev. Lett.}\ }\textbf
		{\bibinfo {volume} {118}},\ \bibinfo {pages} {037601} (\bibinfo {year}
		{2017})}\BibitemShut {NoStop}%
	\bibitem [{\citenamefont {Gulka}\ \emph {et~al.}(2017)\citenamefont {Gulka},
		\citenamefont {Bourgeois}, \citenamefont {Hruby}, \citenamefont {Siyushev},
		\citenamefont {Wachter}, \citenamefont {Aumayr}, \citenamefont {Hemmer},
		\citenamefont {Gali}, \citenamefont {Jelezko}, \citenamefont {Trupke},\ and\
		\citenamefont {Nesladek}}]{GulkaPRAppl17}%
	\BibitemOpen
	\bibfield  {author} {\bibinfo {author} {\bibfnamefont {M.}~\bibnamefont
			{Gulka}}, \bibinfo {author} {\bibfnamefont {E.}~\bibnamefont {Bourgeois}},
		\bibinfo {author} {\bibfnamefont {J.}~\bibnamefont {Hruby}}, \bibinfo
		{author} {\bibfnamefont {P.}~\bibnamefont {Siyushev}}, \bibinfo {author}
		{\bibfnamefont {G.}~\bibnamefont {Wachter}}, \bibinfo {author} {\bibfnamefont
			{F.}~\bibnamefont {Aumayr}}, \bibinfo {author} {\bibfnamefont {P.~R.}\
			\bibnamefont {Hemmer}}, \bibinfo {author} {\bibfnamefont {A.}~\bibnamefont
			{Gali}}, \bibinfo {author} {\bibfnamefont {F.}~\bibnamefont {Jelezko}},
		\bibinfo {author} {\bibfnamefont {M.}~\bibnamefont {Trupke}},\ and\ \bibinfo
		{author} {\bibfnamefont {M.}~\bibnamefont {Nesladek}},\ }\href@noop {}
	{\bibfield  {journal} {\bibinfo  {journal} {Phys. Rev. Applied}\ }\textbf
		{\bibinfo {volume} {7}},\ \bibinfo {pages} {044032} (\bibinfo {year}
		{2017})}\BibitemShut {NoStop}%
	\bibitem [{\citenamefont {Stich}\ \emph {et~al.}(1996)\citenamefont {Stich},
		\citenamefont {Greulich-Weber},\ and\ \citenamefont {Speath}}]{StichAPL96}%
	\BibitemOpen
	\bibfield  {author} {\bibinfo {author} {\bibfnamefont {B.}~\bibnamefont
			{Stich}}, \bibinfo {author} {\bibfnamefont {S.}~\bibnamefont
			{Greulich-Weber}},\ and\ \bibinfo {author} {\bibfnamefont {J.-M.}\
			\bibnamefont {Speath}},\ }\href@noop {} {\bibfield  {journal} {\bibinfo
			{journal} {Appl. Phys. Lett.}\ }\textbf {\bibinfo {volume} {68}},\ \bibinfo
		{pages} {1102} (\bibinfo {year} {1996})}\BibitemShut {NoStop}%
	\bibitem [{\citenamefont {McCamey}\ \emph {et~al.}(2010)\citenamefont
		{McCamey}, \citenamefont {van Tol}, \citenamefont {Morley},\ and\
		\citenamefont {Boehme}}]{McCameySci10}%
	\BibitemOpen
	\bibfield  {author} {\bibinfo {author} {\bibfnamefont {D.~R.}\ \bibnamefont
			{McCamey}}, \bibinfo {author} {\bibfnamefont {J.}~\bibnamefont {van Tol}},
		\bibinfo {author} {\bibfnamefont {G.~W.}\ \bibnamefont {Morley}},\ and\
		\bibinfo {author} {\bibfnamefont {C.}~\bibnamefont {Boehme}},\ }\href@noop {}
	{\bibfield  {journal} {\bibinfo  {journal} {Science}\ }\textbf {\bibinfo
			{volume} {330}},\ \bibinfo {pages} {1652} (\bibinfo {year}
		{2010})}\BibitemShut {NoStop}%
	\bibitem [{\citenamefont {Hoehne}\ \emph {et~al.}(2011)\citenamefont {Hoehne},
		\citenamefont {Dreher}, \citenamefont {Huebl}, \citenamefont {Stutzmann},\
		and\ \citenamefont {Brandt}}]{HoehnePRL11}%
	\BibitemOpen
	\bibfield  {author} {\bibinfo {author} {\bibfnamefont {F.}~\bibnamefont
			{Hoehne}}, \bibinfo {author} {\bibfnamefont {L.}~\bibnamefont {Dreher}},
		\bibinfo {author} {\bibfnamefont {H.}~\bibnamefont {Huebl}}, \bibinfo
		{author} {\bibfnamefont {M.}~\bibnamefont {Stutzmann}},\ and\ \bibinfo
		{author} {\bibfnamefont {M.~S.}\ \bibnamefont {Brandt}},\ }\href@noop {}
	{\bibfield  {journal} {\bibinfo  {journal} {Phys. Rev. Lett.}\ }\textbf
		{\bibinfo {volume} {106}},\ \bibinfo {pages} {187601} (\bibinfo {year}
		{2011})}\BibitemShut {NoStop}%
	\bibitem [{\citenamefont {McCamey}\ \emph {et~al.}(2012)\citenamefont
		{McCamey}, \citenamefont {Boehme}, \citenamefont {Morley},\ and\
		\citenamefont {van Tol}}]{McCameyPRB12}%
	\BibitemOpen
	\bibfield  {author} {\bibinfo {author} {\bibfnamefont {D.~R.}\ \bibnamefont
			{McCamey}}, \bibinfo {author} {\bibfnamefont {C.}~\bibnamefont {Boehme}},
		\bibinfo {author} {\bibfnamefont {G.~W.}\ \bibnamefont {Morley}},\ and\
		\bibinfo {author} {\bibfnamefont {J.}~\bibnamefont {van Tol}},\ }\href@noop
	{} {\bibfield  {journal} {\bibinfo  {journal} {Phys. Rev. B}\ }\textbf
		{\bibinfo {volume} {85}},\ \bibinfo {pages} {073201} (\bibinfo {year}
		{2012})}\BibitemShut {NoStop}%
	\bibitem [{\citenamefont {Dreher}\ \emph {et~al.}(2012)\citenamefont {Dreher},
		\citenamefont {Hoehne}, \citenamefont {Stutzmann},\ and\ \citenamefont
		{Brandt}}]{DreherPRL12}%
	\BibitemOpen
	\bibfield  {author} {\bibinfo {author} {\bibfnamefont {L.}~\bibnamefont
			{Dreher}}, \bibinfo {author} {\bibfnamefont {F.}~\bibnamefont {Hoehne}},
		\bibinfo {author} {\bibfnamefont {M.}~\bibnamefont {Stutzmann}},\ and\
		\bibinfo {author} {\bibfnamefont {M.~S.}\ \bibnamefont {Brandt}},\
	}\href@noop {} {\bibfield  {journal} {\bibinfo  {journal} {Phys. Rev. Lett.}\
		}\textbf {\bibinfo {volume} {108}},\ \bibinfo {pages} {027602} (\bibinfo
		{year} {2012})}\BibitemShut {NoStop}%
	\bibitem [{\citenamefont {Malissa}\ \emph {et~al.}(2014)\citenamefont
		{Malissa}, \citenamefont {Kavand}, \citenamefont {Waters}, \citenamefont {van
			Schooten}, \citenamefont {Burn}, \citenamefont {Vardeny}, \citenamefont
		{Saam}, \citenamefont {Lupton},\ and\ \citenamefont {Boehme}}]{MalissaSci14}%
	\BibitemOpen
	\bibfield  {author} {\bibinfo {author} {\bibfnamefont {H.}~\bibnamefont
			{Malissa}}, \bibinfo {author} {\bibfnamefont {M.}~\bibnamefont {Kavand}},
		\bibinfo {author} {\bibfnamefont {D.~P.}\ \bibnamefont {Waters}}, \bibinfo
		{author} {\bibfnamefont {K.~J.}\ \bibnamefont {van Schooten}}, \bibinfo
		{author} {\bibfnamefont {P.~L.}\ \bibnamefont {Burn}}, \bibinfo {author}
		{\bibfnamefont {Z.~V.}\ \bibnamefont {Vardeny}}, \bibinfo {author}
		{\bibfnamefont {B.}~\bibnamefont {Saam}}, \bibinfo {author} {\bibfnamefont
			{J.~M.}\ \bibnamefont {Lupton}},\ and\ \bibinfo {author} {\bibfnamefont
			{C.}~\bibnamefont {Boehme}},\ }\href@noop {} {\bibfield  {journal} {\bibinfo
			{journal} {Science}\ }\textbf {\bibinfo {volume} {345}},\ \bibinfo {pages}
		{1487} (\bibinfo {year} {2014})}\BibitemShut {NoStop}%
	\bibitem [{\citenamefont {Boehme}\ and\ \citenamefont
		{Lips}(2003)}]{BoehmePRB03}%
	\BibitemOpen
	\bibfield  {author} {\bibinfo {author} {\bibfnamefont {C.}~\bibnamefont
			{Boehme}}\ and\ \bibinfo {author} {\bibfnamefont {K.}~\bibnamefont {Lips}},\
	}\href@noop {} {\bibfield  {journal} {\bibinfo  {journal} {Phys. Rev. B}\
		}\textbf {\bibinfo {volume} {68}},\ \bibinfo {pages} {245105} (\bibinfo
		{year} {2003})}\BibitemShut {NoStop}%
	\bibitem [{\citenamefont {Doi}\ \emph {et~al.}(2016)\citenamefont {Doi},
		\citenamefont {Fukui}, \citenamefont {Kato}, \citenamefont {Makino},
		\citenamefont {Yamasaki}, \citenamefont {Tashima}, \citenamefont {Morishita},
		\citenamefont {Miwa}, \citenamefont {Jelezko}, \citenamefont {Suzuki},\ and\
		\citenamefont {Mizuochi}}]{DoiPRB2016}%
	\BibitemOpen
	\bibfield  {author} {\bibinfo {author} {\bibfnamefont {Y.}~\bibnamefont
			{Doi}}, \bibinfo {author} {\bibfnamefont {T.}~\bibnamefont {Fukui}}, \bibinfo
		{author} {\bibfnamefont {H.}~\bibnamefont {Kato}}, \bibinfo {author}
		{\bibfnamefont {T.}~\bibnamefont {Makino}}, \bibinfo {author} {\bibfnamefont
			{S.}~\bibnamefont {Yamasaki}}, \bibinfo {author} {\bibfnamefont
			{T.}~\bibnamefont {Tashima}}, \bibinfo {author} {\bibfnamefont
			{H.}~\bibnamefont {Morishita}}, \bibinfo {author} {\bibfnamefont
			{S.}~\bibnamefont {Miwa}}, \bibinfo {author} {\bibfnamefont {F.}~\bibnamefont
			{Jelezko}}, \bibinfo {author} {\bibfnamefont {Y.}~\bibnamefont {Suzuki}},\
		and\ \bibinfo {author} {\bibfnamefont {N.}~\bibnamefont {Mizuochi}},\
	}\href@noop {} {\bibfield  {journal} {\bibinfo  {journal} {Phys. Rev. B}\
		}\textbf {\bibinfo {volume} {93}},\ \bibinfo {pages} {081203(R)} (\bibinfo
		{year} {2016})}\BibitemShut {NoStop}%
	\bibitem [{\citenamefont {Kato}\ \emph {et~al.}(2005)\citenamefont {Kato},
		\citenamefont {Yamasaki},\ and\ \citenamefont {Okushi}}]{KatoAPL05}%
	\BibitemOpen
	\bibfield  {author} {\bibinfo {author} {\bibfnamefont {H.}~\bibnamefont
			{Kato}}, \bibinfo {author} {\bibfnamefont {S.}~\bibnamefont {Yamasaki}},\
		and\ \bibinfo {author} {\bibfnamefont {H.}~\bibnamefont {Okushi}},\
	}\href@noop {} {\bibfield  {journal} {\bibinfo  {journal} {Appl. Phys.
				Lett.}\ }\textbf {\bibinfo {volume} {86}},\ \bibinfo {pages} {222111}
		(\bibinfo {year} {2005})}\BibitemShut {NoStop}%
	\bibitem [{SM()}]{SM}%
	\BibitemOpen
	\href@noop {} {\bibinfo {title} {Supplemental material}}\BibitemShut
	{NoStop}%
	\bibitem [{\citenamefont {Schweiger}\ and\ \citenamefont
		{Jeschke}(2001)}]{pEPRBook}%
	\BibitemOpen
	\bibfield  {author} {\bibinfo {author} {\bibfnamefont {A.}~\bibnamefont
			{Schweiger}}\ and\ \bibinfo {author} {\bibfnamefont {G.}~\bibnamefont
			{Jeschke}},\ }\href@noop {} {\emph {\bibinfo {title} {Principles of pulse
				electron paramagnetic resonance}}}\ (\bibinfo  {publisher} {Oxford Universtiy
		Press, New York},\ \bibinfo {year} {2001})\ Chap.\ \bibinfo {chapter} {10,
		11, and 12}\BibitemShut {NoStop}%
	\bibitem [{\citenamefont {Hoehne}\ \emph {et~al.}(2013)\citenamefont {Hoehne},
		\citenamefont {Dreher}, \citenamefont {Suckert}, \citenamefont {Franke},
		\citenamefont {Stutzmann},\ and\ \citenamefont {Brandt}}]{HoehnePRB13}%
	\BibitemOpen
	\bibfield  {author} {\bibinfo {author} {\bibfnamefont {F.}~\bibnamefont
			{Hoehne}}, \bibinfo {author} {\bibfnamefont {L.}~\bibnamefont {Dreher}},
		\bibinfo {author} {\bibfnamefont {M.}~\bibnamefont {Suckert}}, \bibinfo
		{author} {\bibfnamefont {D.~P.}\ \bibnamefont {Franke}}, \bibinfo {author}
		{\bibfnamefont {M.}~\bibnamefont {Stutzmann}},\ and\ \bibinfo {author}
		{\bibfnamefont {M.~S.}\ \bibnamefont {Brandt}},\ }\href@noop {} {\bibfield
		{journal} {\bibinfo  {journal} {Phys. Rev. B}\ }\textbf {\bibinfo {volume}
			{88}},\ \bibinfo {pages} {155301} (\bibinfo {year} {2013})}\BibitemShut
	{NoStop}%
	\bibitem [{\citenamefont {He}\ \emph {et~al.}(1993)\citenamefont {He},
		\citenamefont {Manson},\ and\ \citenamefont {Fisk}}]{HePRB93}%
	\BibitemOpen
	\bibfield  {author} {\bibinfo {author} {\bibfnamefont {X.-F.}\ \bibnamefont
			{He}}, \bibinfo {author} {\bibfnamefont {N.~B.}\ \bibnamefont {Manson}},\
		and\ \bibinfo {author} {\bibfnamefont {P.~T.~H.}\ \bibnamefont {Fisk}},\
	}\href@noop {} {\bibfield  {journal} {\bibinfo  {journal} {Phys. Rev. B}\
		}\textbf {\bibinfo {volume} {47}},\ \bibinfo {pages} {8816} (\bibinfo {year}
		{1993})}\BibitemShut {NoStop}%
	\bibitem [{\citenamefont {Felton}\ \emph {et~al.}(2009)\citenamefont {Felton},
		\citenamefont {Edmonds}, \citenamefont {Newton}, \citenamefont {Martineau},
		\citenamefont {Fisher}, \citenamefont {Twitchen},\ and\ \citenamefont
		{Baker}}]{FeltonPRB09}%
	\BibitemOpen
	\bibfield  {author} {\bibinfo {author} {\bibfnamefont {S.}~\bibnamefont
			{Felton}}, \bibinfo {author} {\bibfnamefont {A.~M.}\ \bibnamefont {Edmonds}},
		\bibinfo {author} {\bibfnamefont {M.~E.}\ \bibnamefont {Newton}}, \bibinfo
		{author} {\bibfnamefont {P.~M.}\ \bibnamefont {Martineau}}, \bibinfo {author}
		{\bibfnamefont {D.}~\bibnamefont {Fisher}}, \bibinfo {author} {\bibfnamefont
			{D.~J.}\ \bibnamefont {Twitchen}},\ and\ \bibinfo {author} {\bibfnamefont
			{J.~M.}\ \bibnamefont {Baker}},\ }\href@noop {} {\bibfield  {journal}
		{\bibinfo  {journal} {Phys. Rev. B}\ }\textbf {\bibinfo {volume} {79}},\
		\bibinfo {pages} {075203} (\bibinfo {year} {2009})}\BibitemShut {NoStop}%
	\bibitem [{\citenamefont {Yavkin}\ \emph {et~al.}(2016)\citenamefont {Yavkin},
		\citenamefont {Mamin},\ and\ \citenamefont {Orlinskii}}]{YavkinJMR16}%
	\BibitemOpen
	\bibfield  {author} {\bibinfo {author} {\bibfnamefont {B.~V.}\ \bibnamefont
			{Yavkin}}, \bibinfo {author} {\bibfnamefont {G.~V.}\ \bibnamefont {Mamin}},\
		and\ \bibinfo {author} {\bibfnamefont {S.~B.}\ \bibnamefont {Orlinskii}},\
	}\href@noop {} {\bibfield  {journal} {\bibinfo  {journal} {J. Magn. Reson.}\
		}\textbf {\bibinfo {volume} {262}},\ \bibinfo {pages} {15} (\bibinfo {year}
		{2016})}\BibitemShut {NoStop}%
	\bibitem [{\citenamefont {Myers}\ \emph {et~al.}(2017)\citenamefont {Myers},
		\citenamefont {Ariyaratne},\ and\ \citenamefont {Jayich}}]{MyersPRL17}%
	\BibitemOpen
	\bibfield  {author} {\bibinfo {author} {\bibfnamefont {B.~A.}\ \bibnamefont
			{Myers}}, \bibinfo {author} {\bibfnamefont {A.}~\bibnamefont {Ariyaratne}},\
		and\ \bibinfo {author} {\bibfnamefont {A.~C.~B.}\ \bibnamefont {Jayich}},\
	}\href@noop {} {\bibfield  {journal} {\bibinfo  {journal} {Phys. Rev. Lett.}\
		}\textbf {\bibinfo {volume} {118}},\ \bibinfo {pages} {197201} (\bibinfo
		{year} {2017})}\BibitemShut {NoStop}%
	\bibitem [{\citenamefont {Ariyaratne}\ \emph {et~al.}(2018)\citenamefont
		{Ariyaratne}, \citenamefont {Bluvstein}, \citenamefont {Myers},\ and\
		\citenamefont {Jayich}}]{AriyaratneNC18}%
	\BibitemOpen
	\bibfield  {author} {\bibinfo {author} {\bibfnamefont {A.}~\bibnamefont
			{Ariyaratne}}, \bibinfo {author} {\bibfnamefont {D.}~\bibnamefont
			{Bluvstein}}, \bibinfo {author} {\bibfnamefont {B.~A.}\ \bibnamefont
			{Myers}},\ and\ \bibinfo {author} {\bibfnamefont {A.~C.~B.}\ \bibnamefont
			{Jayich}},\ }\href@noop {} {\bibfield  {journal} {\bibinfo  {journal} {Nat.
				Commun.}\ }\textbf {\bibinfo {volume} {10}},\ \bibinfo {pages} {1038}
		(\bibinfo {year} {2018})}\BibitemShut {NoStop}%
\end{thebibliography}

\begin{thebibliography}{2}
	\bibitem[S1]{SDoherty13} M. W. Doherty, N. B. Manson, P. Delaney, F. Jelezko, J. Wrachtrup, and L. C. L. Hollenberg, Phys. Rep. \textbf{528}, 1 (2013).
	\bibitem[S2]{SMackRMP50} J. E. Mack, Rev. Mod. Phys. \textbf{22}, 64 (1950).
	\bibitem[S3]{SFeltonPRB09} S. Felton, A. M. Edmonds, M. E. Newton, P. M. Martineau, D. Fisher, D. J. Twitchen, and J. M. Baker, Phys. Rev. B \textbf{79}, 075203 (2009).
	\bibitem[S4]{SCFMBook}M. M\"{u}ller, \textit{Introduction to Confocal Fluorescence Microscopy, Second Edition} (SPIE The International Society for Optical Engineering, Washington, 2006) Chap. 1.
	\bibitem[S5]{SpEPRBook}A. Schweiger and G. Jeschke, \textit{Principles of pulse electron paramagnetic resonance} (Oxford Universtiy Press, New York, 2001) Chap. 10, 11, and 12.
	\bibitem[S6]{SHoehnePRB13}F. Hoehne, L. Dreher, M. Suckert, D. P. Franke, M. Stutzmann, and M. S. Brandt, Phys. Rev. B \textbf{88}, 155301 (2013).
	\bibitem[S7]{SMalissaSci14} H. Malissa, M. Kavand, D. P. Waters, K. J. van Schooten, P. L. Burn, Z. V. Vardeny, B. Saam, J. M. Lupton, and C. Boehme, Science \textbf{345}, 1487 (2014).
\end{thebibliography}
\end{document}